\documentclass[journal]{IEEEtran}

\ifCLASSINFOpdf
   \else
  
\fi

\usepackage{cite}
\usepackage{graphicx}
\usepackage{float}
\usepackage{ifpdf}
\usepackage{fixltx2e}
\usepackage{caption}
 \usepackage{url}  
 \usepackage{hyperref}
 \usepackage{lipsum}
 \usepackage{xcolor}
\usepackage[fleqn]{amsmath}
\usepackage{array}
\usepackage{booktabs}
\usepackage{float}
\usepackage{fancyhdr}
\usepackage{mathtools}

\usepackage{multirow}
\usepackage{lscape}
\usepackage{ragged2e,array,booktabs}
 \usepackage{amsfonts}
\usepackage{pdflscape}
\usepackage{rotating}
\usepackage[utf8]{inputenc}
\usepackage[english]{babel}
\usepackage{soul}

\begin{document}

	\title{Authentication of Smartphone Users Using Behavioral Biometrics}
	
	\author{\IEEEauthorblockN{Abdulaziz Alzubaidi and Jugal Kalita}
	}


	\maketitle

\begin{abstract}
Smartphones and tablets have become ubiquitous in our daily lives. Smartphones, in particular, have become more than personal assistants. These devices have provided new avenues for consumers to play, work and socialize whenever and wherever they want. Smartphones are small in size; so they are easy to handle and to stow and carry in users' pockets or purses. However, mobile devices are also susceptible to various problems. One of the greatest concerns is the possibility of breach in security and privacy if the device is seized by an outside party. It is possible that threats can come from friends as well as strangers. Due to the size of smart devices, they can be easily lost and may expose details of users' private lives. In addition, this might enable pervasive observation or imitation of one's movements and activities, such as sending messages to contacts, accessing private communication, shopping with a credit card, and relaying information about where one has been. This paper highlights the potential risks that occur when smartphones are stolen or seized, discusses the concept of continuous authentication, and analyzes current approaches and mechanisms of behavioral biometrics with respect to methodology, associated datasets and evaluation approaches.
\end{abstract}


\IEEEpeerreviewmaketitle

\section{Introduction}

\IEEEPARstart{O}{ver} the last few years, the world has witnessed the beginning of a revolution taking shape in the field of technology. One of the greatest innovations in technology is the smartphone device. Smartphone devices are characterized by expedient features, such as sophisticated operating systems that can allow users to browse the Internet; to listen, watch, and record video streams; and to navigate, using GPS. These devices also have large internal storage that enables users to store gigabytes of valuable information, such as personal photos, contact details, call histories and private messages. Rapid progress in mobile technology has led to a significant shift in large numbers of consumers using smartphone devices instead of personal computers. Market research finds that the number of smartphones sold has surpassed the number of laptops sold worldwide \cite{bworld}. The tremendous increase in the number of consumers who are buying smartphones has pushed these devices to the top of the market, and they now lead all other electronic devices in terms of sales. According to  the International Data Corporation (IDC), the total number of shipments in the second quarter of 2015 reached 337.2 million smartphones worldwide, an increase of 11.6\% compared to the same quarter in 2014 \cite{IDS1}. The second quarter in 2015 has the second highest quarterly total on record.  The number of smartphones is shipped predicted to rise to 1,928.4 million in 2019 \cite{IDS2}. \\

As much as these devices have gained in popularity and enhanced users' productivity and consumption of entertainment, the security of these devices continues to be a major concern for manufacturers and users alike. This paper focuses on current methods for user authentication on smartphone devices. 

\subsection*{Motivation}

In this paper, we plan to comprehensively review the state-of-the-art in smartphone authentication focusing on seven types of behavioral biometrics, which are handwaving, gait, touchscreen, keystroke, voice, signature and general profiling. We summarize existing studies which propose significant solutions for smartphone authentication by discussing the following points:
\begin{itemize}	
	\item the amount of the data the authors use,
	\item the types of classifiers the authors choose, and 
	\item the results the authors obtain.
\end{itemize}

Our discussions are extensive and we follow by outlining lessons learned and a list of open problems so that new researchers can get started quickly and avoid pitfalls in their work. Obviously, we focus on potential solutions that can be installed on the smartphone platform.

\subsection*{Why this survey}

Research on authentication of smartphone users has seen concentrated work during the last few years, many based on behavioral biometrics, looking at user behavior concerning the use of keystrokes and touchscreens, and the individual characteristics of  handwaving and gait. With many recent studies in the fields, there is need for a survey of techniques that have been published. Compared to recent published surveys such as \cite{ashbourn2014biometrics}, \cite{long2014biometrics}, \cite{goode2014bring} and \cite{meng2015surveying} on the authentication of smartphone users based on behavioral biometrics, our survey differs in the following ways.\\

\begin{itemize}	
	\item Unlike Meng et al. \cite{meng2015surveying}, we extensively cover seven behavioral biometrics in terms of methodology, associated datasets and evaluation approaches. This survey offers a good background for any researcher who is interested in this field.  
	
	\item We investigate the awarenesses of security needs that users express based on surveys. Having a good understanding of users' perceived and real needs and practices is important  so that researchers are well- informed about what approaches will be accepted by users and what will be ignored. 
	
\end{itemize}

Table \ref{table_CSS} shows the comparison between our survey and the most current surveys in this field.\\

\begin{table*}[htp]
	\renewcommand{\arraystretch}{1.3}
	\caption{Comparison with Other surveys, {where \checkmark: cover the topic, x: does not cover the topic.}}
	\label{table_CSS}
	\centering
	\renewcommand{\arraystretch}{1.2}
	\begin{tabular}{|p{3.5cm}|p{2.5cm}|p{0.5cm}|p{1.4cm}|p{1.1cm}|p{1cm}|p{1.5cm}|p{0.5cm}|p{1.3cm}|}
		\hline
		\textbf{Study} & \multicolumn{8}{c|}{\textbf{Behavioral Biometric}}\\
		
		\cline{2-9}
		& \textbf{Behavioral Profiling} & \textbf{Gait} & \textbf{Handwaving} & \textbf{Keystroke} & \textbf{Signature} & \textbf{Touchscreen}& \textbf{Voice} & \textbf{User view}  \\
		\hline\hline
		
		Crawford \cite{crawford2010keystroke}, 2010 & x & x& x & \checkmark & x & x & x&x\\
		\hline 
		
		Duta \cite{duta2009survey}, 2009 & x&x&x&x&x&x&x&x\\
		\hline 
		
		Yampolskiy and Govindaraju \cite{yampolskiy2008behavioural}, 2008 &\checkmark&x&x&x&x&x&x&x\\
		\hline 
		
		Zhang and Gao \cite{zhang2009face}, 2009 & x&x&x&x&x&x&x&x\\
		\hline 
		
		Ashbourn\cite{ashbourn2014biometrics}, 2014 &x&x&x&\checkmark&\checkmark&x&\checkmark&x\\
		\hline 
		
		Long\cite{jain2007biometric}, 2007  & x&x&x&x&x&x&\checkmark&x\\
		\hline

		Jain \cite{long2014biometrics}, 2014 &x&\checkmark&x&\checkmark&\checkmark&x&\checkmark&x\\
		\hline

		Gooda \cite{goode2014bring}, 2014 &x&x&x&x&x&x&\checkmark&x\\
		\hline

		Rogowski et al.'s \cite{rogowski2013user}, 2013 &x&\checkmark&x&\checkmark&x&\checkmark&\checkmark&x\\
		\hline

		Meng et al.\cite{meng2015surveying}, 2015 &\checkmark&\checkmark&x&\checkmark&\checkmark&\checkmark&\checkmark&x\\
		\hline

		Zada Khan et al. \cite{khan2013mobile}, 2013 &x&x&x&x&x&x&x&x\\
		\hline

		Hoseini-Tabatabaei et al.\cite{hoseini2013survey}, 2013 &x&x&x&x&x&x&x&x\\
		\hline
		
		Our Survey, 2015 & \checkmark&\checkmark&\checkmark&\checkmark&\checkmark&\checkmark&\checkmark&\checkmark\\ 
		
		\hline
		
	\end{tabular}
	
\end{table*}

\subsection*{\textbf{Contribution and Summary}}

The scope of this survey is continuous authentication for smartphones platforms. Our contributions in this paper are as follows.

\begin{itemize}
	\item We discuss the development of several behavior biometric approaches that aim to provide continuous authentication for smartphone devices.
	
	\item We characterize each behavioral biometric, outline the algorithms used for recognition and  present obtaining results obtained using various techniques for comparison.
	
	\item We present a summary of these studies and introduce open problems and future work in continuous authentication.\\
\end{itemize}

The remainder of this paper is organized as follows. Section \ref{sec:P&S} discusses possible threats that could impact upon security and privacy for smartphone consumers and  outlines characteristics of authentication solutions. Section \ref{sec:CONC} introduces basic concepts that are related to the research topic, i.e., biometrics, authentication, and smartphone sensors. Section \ref{sec:MLEM} outlines evaluation metrics used in most approaches. Types of behavioral biometrics used in continuous authentication, collection data, classifiers used and results are discussed in Section \ref{sec:A2A}. Lessons learned and open problems are presented in Section \ref{sec:LLOP}.  

\section{PROBLEM \& Solution Characteristics}
\label{sec:P&S}

In this section, we introduce the problem we focus on and general characteristics of solutions proposed by researchers in this field.

\subsection{Problem}
\label{sec:PROBLEM}
Smartphones provide substantial personal benefits for users every day. Many people have come to rely on smartphones for many common, personal and work-related communication tasks. Most users tend to store their passwords and private information on smartphones to efficiently perform these operations in a hassle-free manner. Consequently, potential threats to the accounts of owners have increased tremendously.\\

With the vast popularity of smartphones, privacy and security issues have become paramount. Due to the size of smartphones, they are quite prone to potentially being lost, stolen, or accessed easily by non-owners. Once an intruder has physical access to a device, he/she may be able to impersonate the original owner of the device for monetary or non-monetary gains and mischiefs; thus, smartphones are much more susceptible to theft than desktops \cite{LAFC}. Attackers are likely to access Online Social Networks (OSNs), financial application and other applications on stolen devices. According to \footnote{http://www.consumerreports.org/cro/news/2014/04/smart-phone-thefts-rose-to-3-1-million-last-year/index.htm}\cite{STR}, the total number of lost or stolen devices in the USA increased from 1.6 million in 2012 to 3.1 million in 2013. Breitinger and Nickel's survey\cite{breitinger2010user} of 548 subjects shows that only 13\% of owners tends to use PIN or visual codes, which means that information contained in the smartphones of at least 87\% of the owners is in danger once these devices are lost or stolen. 74\% of the participants justify this by saying that they want quick access to their devices or that they do not think about security.\\

As a result, protecting the security and privacy of smartphone users against unauthorized access is very important and has become a crucial area of research. Researchers from both academia and industry have proposed mechanisms to ensure security and privacy of sensitive information. Each one of these mechanisms has strengths as well as drawbacks. Some of the weaknesses of the current security mechanisms are that they are easy to evade, weak against shoulder surfing and other attacks and cumbersome to use \cite{suo2005graphical}. To illustrate, a locking mechanism (e.g., PIN, biometric or secret gesture) supposedly thwarts most potential threats. But, according to Li et al. \cite{li2013unobservable} more than 30\% of mobile phone users do not even use a PIN on their phones. In addition, users' continuous and casual interaction with these devices can make breaching these PINs easier. A smart intruder can, with some diligence, monitor and measure users' behaviors over a period of time to break the authentication and steal passwords.

\subsection{SOLUTION CHARACTERISTICS}
\label{sec:SOLUTION}
\begin{figure}[!h]
	\centering
	\includegraphics[width=3.5in]{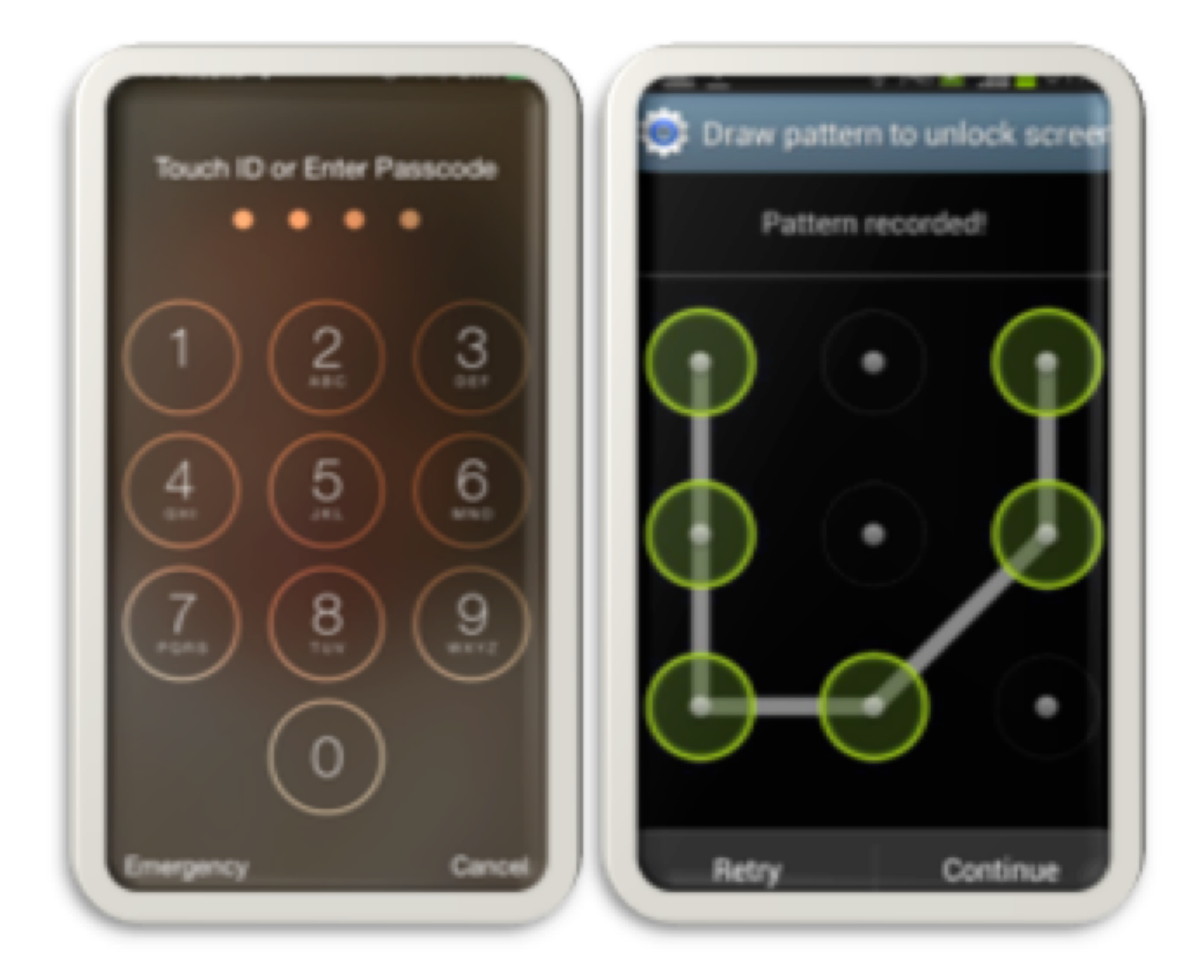}
	\caption{Examples of Smartphone  Entry Authentication, Left: Entering a Passcode, Right: Drawing a Pattern}
	\label{fig_secure}
\end{figure}

Unfortunately, most widely-used authentication techniques for mobile devices are vulnerable, including PINs and patterns as shown in Fig.\ref{fig_secure}. Indeed, these authentication methods fail to detect and identify an intruder once he or she has passed the point of entry. These methods are also deficient in dealing with various non-conventional attacks such as smudge attack \cite{smudge}, which picks up oils from users' skin to detect patterns or PINs, or a make conventional one such as the shoulder surfing attack, which uses direct observation techniques like glancing over the shoulder of a user to gain information\cite{suo2005graphical}. So, what is the most viable solution to these security problems? An obvious solution is to perform entry-point authentication as usual, but go beyond it by performing authentication constantly as a user uses the device. Currently, there are two ways to perform continuous authentication: using physiological and behavioral biometrics. Physiological biometric authentication relies on user's static physical attributes such as fingerprints, facial features or retina images, whereas behavioral biometric authentication adapts to identify features of user's behavior that do not vary over a period of time during daily activities such as typing motions, photo manipulations and hand motions.

A viable authentication solution should be continuous and be able to implicitly confirm the correct identity of the user. The purpose of such an approach is to monitor interactions between a user and the device to decide if the current user actually owns the device or if the device has been lost or stolen and used improperly. As researchers deliberate among current authentication options, we must consider how we can implement a solution that is fast, convenient and easy to use. A viable solution must be able to resolve existing issues, such as security attacks (e.g. smudge attacks \cite{smudge} and shoulder surfing attacks \cite{suo2005graphical}), overcome evasions, and also be practical. The method must also be able to perform passive and continuous validation to create tighter security.

\section{Background Concepts}
\label{sec:CONC}
Methods that work to enhance security and privacy need to pay particular attention to authentication. This section introduces concepts that are necessary to discuss methods of information security, in particular authentication. 

\subsection{Authentication}
Authentication is the process used to validate the true user of a system. Authentication, in the context of security, takes into account three primary strategies\cite{burr2004electronic}. 
\begin{figure}[H]
	\centering
	\includegraphics[width=3.5in]{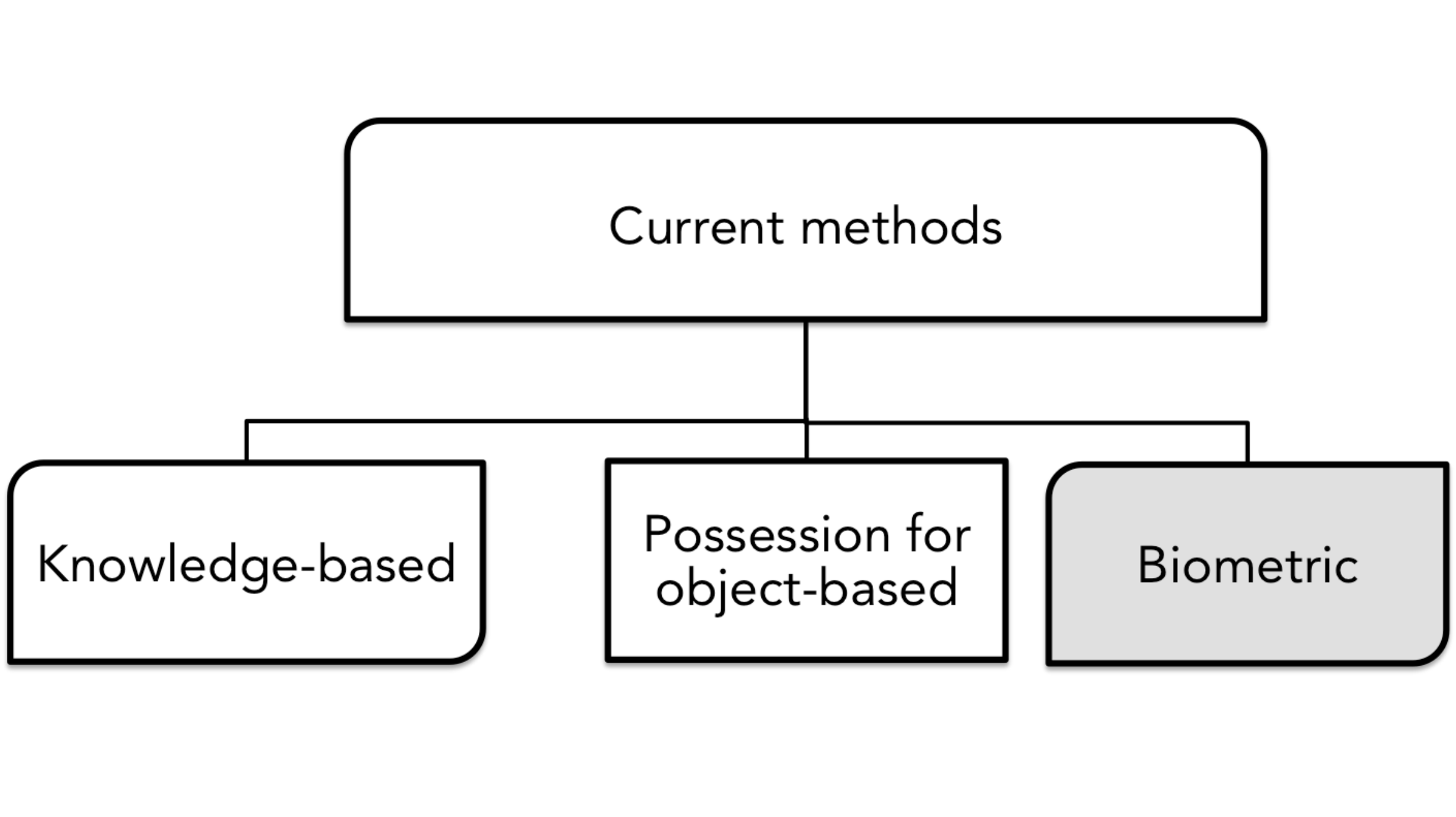}
	\caption{Authentication Strategies for Interaction between Users and Smart devices}\label{Fig_WS}
\end{figure}

\begin{enumerate}[\IEEEsetlabelwidth{4)}]
	\item Knowledge-based, which uses something unique to an individual: This type of entity could be a password, answer to a  security question, or an ID number that a user must know.
	\item Possession for object-based, which uses something one possesses in a physical sense: The prevalent examples of this type are a security token, an ID card or another trusted device.
	\item Biometric, which denotes a physical or behavioral characteristic: This can be represented by one or more physical or behavioral attributes. Common examples are fingerprints and keystroke dynamic models of the owner of the device.\\
\end{enumerate}

Figure. \ref{Fig_WS} illustrates these strategies that aim to improve authentication between users and smartphones.

Authentication can be active or passive. Active Authentication requires dealing with a device and inputting one or more pieces of valid information or answers to questions. Because most individuals use many applications or services, this kind of authentication can become tedious and frustrating; if required by individual applications. As a result, many individuals prefer to use their applications with fewer security impediments each time they decide to access their devices \cite{draffinkeysens}. Current smartphones are based on entry-point authentication, which can be either a Personal Index Number (PIN) or a secret pattern. To use PIN, a user is usually required to pick 4 digits for validation. The user has to input this code correctly otherwise, he/she cannot pass the entry-point to his/her device. Another current method is the use of a secret gesture. A secret gesture is defined by moving a finger over the screen to create a certain pattern. This pattern can be used as authentication to allow the user to enter the device.\\

Continuous authentication, also known implicit, passive or progressive authentication, aims to offer another way to prevent unauthorized accesses of smartphones \cite{frank2013touchalytics}. This method works passively in the background of the device to make a decision. It is divided into two phases. First, the user accesses his/her device as usual, but the system records appropriate features as the user goes about his/her business. In the case of touchscreen analysis, the recorded features may include finger movement, speed, X and Y coordinates of fingers and the pressure applied at sampled time points. After observing the user behavior for a period of time, the system learns characteristics of behavior data by performing statistical analysis or using machine learning. Second, at a later time after the user logs in to his/her device by using for example, a PIN, the system continuously compares current user behavior with the learned user model or computed statistical profile to make an authentication decision \cite{li2013unobservable}.\\ 

Any proposed approach based on continuous authentication should include three features: It should be continuous, should not intrude on normal user behavior and should be light-weight. In other words, the system should allow a legitimate user to use the device without interrupting him/her, e.g., by asking for authentication information such as a  PIN each time he/she wants to access the device. In addition,  the authentication approach should use low computational resources \cite{li2013unobservable}.\\

\subsection{Biometric}
\label{sub:Bio}
\begin{figure}[H]
	\centering
	\includegraphics[width=3.5in]{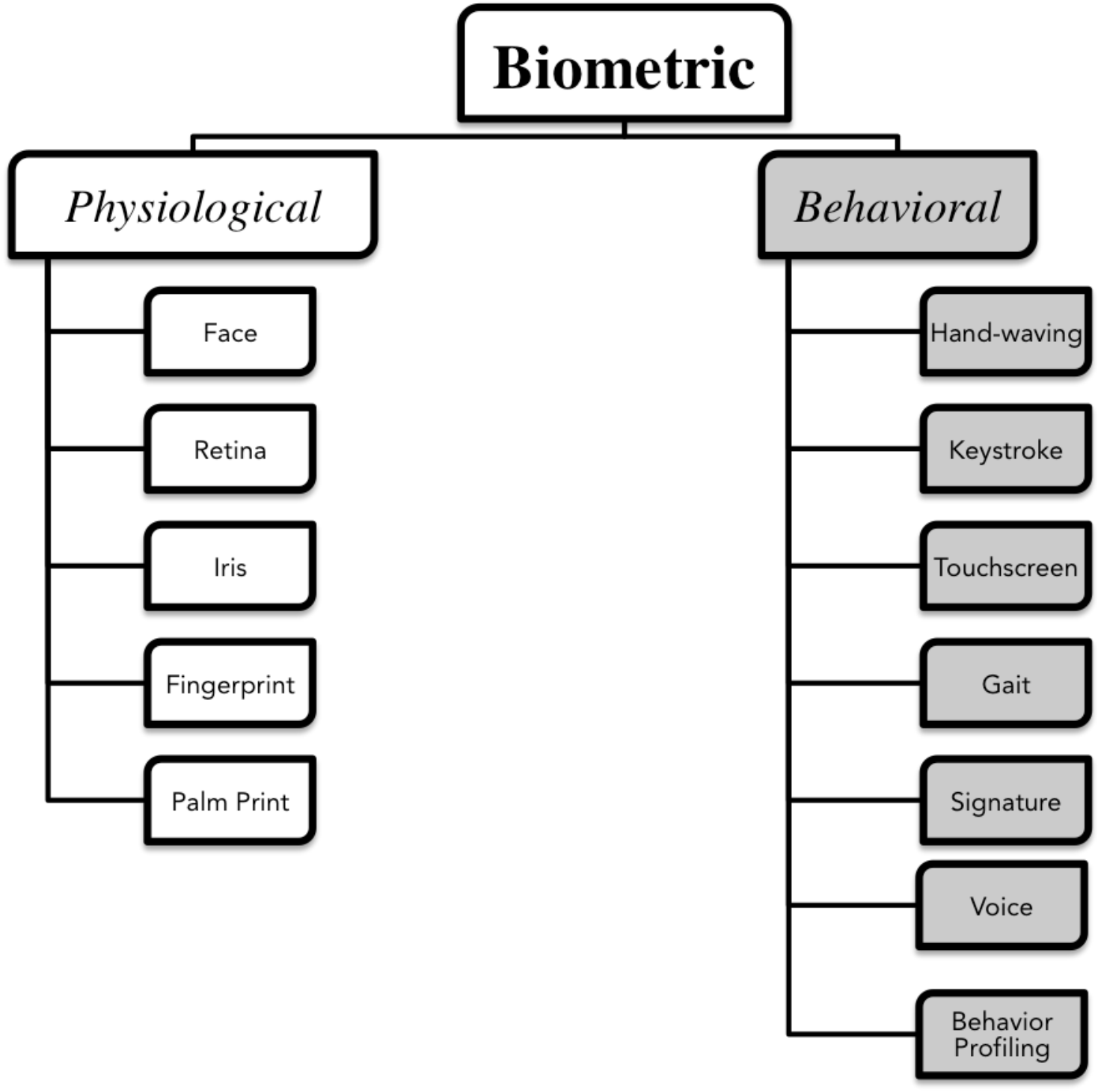}
	\caption{Approaches to Authenticate Users}
	\label{Fig_BS}
\end{figure}

A biometric characterizes unique physical or behavioral features of an individual. A biometric scheme aims to detect and correctly identify the user \cite{burr2004electronic}. The US National Science and Technology Council's Subcommittee (NSTC) on Biometrics defines two categories of biometrics: behavioral and physiological \cite{BIONSTC}. Physiological security detects the physical make-up of a person and uses resources such as retina or iris scans, fingerprints and face recognition. Behavioral biometrics are based on a user's behavior and includes analysis of information like the shape and flow of one's handwriting, timing of keystrokes, unique patterns inherent in one's gait, speech and usage of styluses, and other features of one's general behavior \cite{banerjee2012biometric}. The use of biometrics, generally, is divided into two groups based on the type of application: identification and authentication \cite{banerjee2012biometric}. Fig. \ref{Fig_BS} outlines the two types of biometrics and examples of each type. Once we combine more than one primary type of authentication, we may be able to develop stronger security systems \cite{burr2004electronic}. 

\subsection{Behavior Biometric}
\label{sec:BB}

In this paper, we focus on comprehensively summarizing the state-of-the-art in improving a smartphone's security based on continuous authentication using behavioral biometrics. Behavioral biometrics, as defined in \ref{sub:Bio}, use  behavioral traits of a subject like how one touches screen, walks, talks, signs a signature, and types to identify a subject. Each subject is expected to differ from all others when analyzed using one or more of these features. In the following sections, we discuss in depth four types: keystroke, touchscreen behavior, gait and handwaving, and also introduce other types such as voice, signature and profiling.  A powerful argument for behavioral biometrics is that it can assist in continuous and passive authentication without requiring additional hardware. As a result, behavioral authentication is likely to be cheaper than using physiological biometrics. In the following sections, we will discuss several examples of behavioral biometrics. These are based on touchscreen behavior, gait, keystroke, handwaving, voice, profiling and signature.    

\subsection{Evaluation Metrics}
\label{sec:MLEM}

Various metrics have been used for evaluating each of the proposed mobile authentication systems. For example, to assess the performance of a biometric system, we need to understand two error rates, False Acceptance Rate (FAR) and False Rejection Rate (FRR). Table \ref{table_metric} contains several metrics that have been used to assess implemented approaches. As an example, FAR is defined as the percentage of impersonators who were incorrectly identified by security systems as genuine users\cite{banerjee2012biometric}.\\

\begin{table*}[htp]
	\renewcommand{\arraystretch}{1.3}
	\caption{Common Metrics use for Evaluation of Mobile Authentication}
	\label{table_metric}
	\centering
	\begin{tabular}{|p{3.5cm}||c||p{11cm}|}
		\hline
		\bfseries Concept & \bfseries Acronym & \bfseries Definition \\
		\hline\hline
		False Acceptance Rate & FAR & Proportion of times person are authenticated as rightful owners when they are actually attackers\\
		\hline 
		False Rejection Rate & FRR & Proportion of times the rightful owner is rejected as intruder\\
		\hline 
		True Positive Rate & TPR & Proportion of times when the approach correctly identifies the owner\\
		\hline 
		False Positive Rate & FPR & The likelihood of falsely rejecting the invalid hypothesis for a certain test. \\
		\hline 
		True Negative Rate& TNR & Also known as (Specificity) TNR=TN/(FP+TN)\\
		\hline 
		Receiver Operating Characteristic Curve & ROC & Graphical plot to represent performance based on TRR and FPR \\
		\hline
		Area Under the Curve & AUC & Area under the ROC which is between 0 and 1 \\
		\hline	
		Equal Error Rate & EER & Rate at which both acceptance and rejection errors are closest to zero\\
		\hline 
		Failure to Capture Rate & FCR & Probability that the system fails to detect a biometric input when presented correctly\\
		\hline 
		False Match Rate & FMR & Likelihood of the system falsely matches input to a non-matching value\\
		\hline 
		
		False Non-Match Rate & FNMR & Likelihood of a failure to examine any match between input value and target template \\
		\hline 
		Half Total Error Rate & HTER &  (FMR + FNMR) /2\\
		\hline 
		Total Error Rate & TER &  sum of FAR and FRR\\
		\hline 
		Rectilinear Distance & L1 distance &  $d_1(a,b)$ = $\sum_{i=1}^{n}$ $\mid ai - bi \mid$  \\
		\hline 
		Euclidean distance & L2 distance & d (a,b) = $\sqrt {(a-b).(a-b)} $  \\
		\hline 
		Normalized Cross Correlation & NCC & $\Biggr\langle \frac{A}{||A||}, \frac{B}{||B||} \Biggr\rangle $ \\
		\hline
	\end{tabular}
\end{table*}
\lipsum[0-0]

The purpose of evaluating an authentication system should include the following objectives: providing a trade-off between strong protection and appropriateness of the authentication process, achieving reasonable accuracy in estimating the level of user authenticity, and providing for acceptable execution time and power consumption, i.e., reducing authentication overhead on  smartphone devices.\\

\subsection{Smartphone Sensors}

Most modern smartphones have built-in sensors which can measure motion and environmental and positional environment the devices are subject to. They provide several facilities such as providing accurate and precise raw data, observing the position in three dimensions, and measuring any possible changes in the surrounding environment sufficiently close to the device \cite{MSNS}. The raw measurements may be aggregated by programs or applications to recognize aspects of how people walk, drive, sit up or talk. Many studies in different fields in physiological and behavioral biometrics are based on different sensors to record and extract user's features like the orientation of the device, the pressure on the touchscreen, the pattern of holding the device and the speed of movement. \\

In general, smartphones these days include Android, Apple and Windows platforms and come with three types of sensors, which are Position sensors, Motion sensors and Environmental sensors \cite{MSNS}, \cite{WSENSOR}, \cite{ASENSOR}.\\ 

Position sensors are employed to find the physical position of a device. This group includes several sensors including orientation  sensors and magnetometers. The magnetometer is used to measure the strength and the direction of earth's magnetic fields, which are expressed in tesla. It can be used as a compass, which can be used to find directions in a map. Accelerometers, gyroscopes and magnetometers consistently return three-dimensional values, which are {X or -X}, {Y or -Y} and {Z or -Z} \cite{MSNS} \cite{sen2013study}.\\

Motion sensors measure acceleration and rotational forces along three axes. Examples of such sensors are accelerometers, gravity sensors, gyroscopes and rotational vector sensors \cite{rogowski2013user}. An accelerometer can measure any movement of the phone including fall of the owner when holding the phone or free fall of the phone. A gyroscope detects the current orientation of the device and any possible spin or rotational change. Accelerometers and gyroscopes always return three-dimensional values \cite{rogowski2013user}. The orientation of the smartphone can be computed from the angular velocity detected by the gyroscope, which is expressed on 3 axes
\begin{equation}
S(t) = \overrightarrow{S_x} + \overrightarrow{S_y} + \overrightarrow{S_z} 
\end{equation}
where S(t) denotes the angular velocity, and $\overrightarrow{S_x}$, $\overrightarrow{S_y}$, and $\overrightarrow{S_z}$ refer to the vectors of angular velocity around X, Y, and Z axes \cite{sen2013study}.\\

Environmental sensors measure parameters of the environment. Tools in this category of sensors include barometers, photometers and thermometers \cite{MSNS}\\

Besides these sensors, smartphones also include other sensors such as microphones, cameras, touchscreens, Global Positioning Systems (GPS), and compasses.\\

\section{APPROACHES TO AUTHENTICATION} 
\label{sec:A2A}
This section discusses current research on reinforcing interactions between users and smart devices so that authentication becomes seamless. Studies show that the security and privacy of smartphones and smart devices can be implemented better by using implicit or continuous authentication. Continuous or implicit authentication can potentially offer a stronger line of defense and implement passive security with non-intrusive measures. Such approaches analyze interactions of users with devices and build approximate models of situations when legal users are the ones using the devices. A continuous authentication strategy can strike a balance between passive authentication and entry-point authentication based on the examination of successful logins. In certain situations, a security method may even be able to substitute active authentication with continuous authentication. This section discusses several studies that aim to enhance security based on continuous authentication. These studies use different methods like environment authentication, typing and touchscreen-based authentication, authentication based on user activities and authentication based on gait. Fig. \ref{Fig_BS} outlines these approaches.  

\subsection{Handwaving Based Authentication}
Identifying users based on wave gesture has gained attention recently. Hand-waving behavior is the waving pattern of a person. In other words, it can be used to distinguish users because different individual, while interacting with the phone or not, the movement of hand holding the phone is difficult for different people wave differently. For example, many people use their hands to wave in a gentle way while  others wave drastically when an individual waves while holding a smartphone. Several features can be used to distinguish among users. These include speed, frequency, waving range and the wrist twisting. \\

\subsubsection{Wave-to-Access}

Shrestha et al. \cite{shrestha2013wave} introduced an approach  called Wave-to-Access based on waving gestures to prevent malware attack on smartphones. This approach uses a lightweight ambient light sensor that is built in smartphones, to analyze phone dialing behavior. The authorized user has to wave his/her hand in front of the phone order to make a call. The authors implemented an application on Motorola Droid X2 using Android OS. The authors recruited 20 subjects, and each subject had to initially perform hand waving 10 times to unlock the device and make phone calls. The authors evaluated this experiment in terms of FNR and the average of FNR obtained was 9.5\%. The second experiment calculated FPR when devices were used as several activities are performed concurrently. The authors simulated activities such as walking while phone is inside a backpack, walking and reading a text message, walking and keeping the phone inside the pocket, going upstairs/downstairs, placing the phone in front of the TV while watching movies and playing a game on one's own phone. The best FPR achieved in each case was between 0.08\%-1.15\%.\\

\subsubsection{OpenSesame}

Yang et al.\cite{yang2015unlocking} implemented an approach, called OpenSesame, that employs users' waving patterns for locking/unlocking purposes. The authors asked 200 subjects to participate in this study. 389,373 raw tuples are captured with an average 1,947 tuples per user. Each user has to shake his/her smartphone for approximately 10 seconds for four times. There were two modes, normal and fast. In the normal mode, a user's traces consisted of 100 samples at 200 ms intervals, while in the fast mode, the user's traces consisted of 100 samples with the accelerometer sampling every 10-20 ms. The authors generated two features, which were window size and tolerant static period that represents by acceleration point. The window size \textit{w} has to be smaller than the size of the dataset (1,000 acceleration points).\\

The authors used support vector machines (SVM) \cite{SVMs} for classifying users into two classes: authorized and unauthorized. The authors use the open source library, LIB-SVM \cite{chang2011libsvm}. The result of this approach can be presented from three perspectives: Impact of SVM, Impact of user motion and usability.\\

First, the Impact of SVM: In order to balance between time needed to unlock a device and the accuracy, the window size should be chosen carefully. The authors used window size of 50 instead of 5 tuples. They found that the average FNR is reduced from 20\% to 8\%  and the average FPR decreases from 42\% to 18\% because a large number of windows can help enhance accuracy since more raw tuples are extracted in a larger window and the user's handwaving is better characterized. More training tuples improved the accuracy as well. The authors found FNR decreases from 15\% to 8\% and FPR reduces from 20\% to 15\% . Second, the Impact of User Motion:
The authors tested the relationship between the speed of user's motions and accuracy. Five types of user motions were considered: stationary, walking slow, walking fast, running, and being in a vehicle. As the speed grows from 0 to 5 ms, the FNR remains is steady around 11\%, with a standard deviation of 2.0\%. This indicates that the motion of users makes a very limited effect on the approach. The FPR is actually invariant when the speed of user's motion increases. The false positive rate is around 15 percent with a standard deviation of 2.5\%. The FNR has a slight increase, about 7\%, when the speed of user increases from 0 to 5 m/s. This can be explained because the faster motion increases vibrations in smartphones leading to more noisy data capture. However, these motions have very limited effect on the accuracy.\\

Finally, the authors computed the required time to unlock the devices. They used HTC One, Xiaomi 2, Nexus 5, Huawei C8815 and Sony Xperia. The time needed to unlock devices was 3 seconds, which is higher than using PIN. This means that this approach does not reduce the time for unlocking devices. The authors also asked the participants to evaluate the approach from different perspectives such as user-friendliness, security and accessibility. Feedback from the participants and results obtained showed this approach is easy to use.\\

Table \ref{table_HW} summarizes current studies that use handwaving biometric.\\

\begin{table}[!htp]
	\renewcommand{\arraystretch}{1.3}
	\caption{Examples of Handwaving}
	\label{table_HW}
	\centering
	\renewcommand{\arraystretch}{1.2}
	\begin{tabular}{|p{1.1cm}|p{.8cm}|p{1.3cm}|p{1.6cm}|p{.8cm}|p{.7cm}|}
		\hline
		\textbf{Study} & \textbf{Dataset} & \textbf{Classifier} & \textbf{Used Feature} & \multicolumn{2}{c|}{\textbf{RESULT}}\\
		
		\cline{5-6}
		&&&&   \textbf{FPR} & \textbf{FNR}  \\
		\hline\hline
		
		Shrestha et al. \cite{shrestha2013wave}, 2013 & 20 & Linear Accelerometer Sensor \cite{chen1994gyroscope} & Timestamps, Light intensity \& Duration for the hand wave gesture& 0.08\%- 1.15\% &  9.5\%\\

		\hline
		
		Yang et al.\cite{yang2015unlocking},2015 & 200 & SVM \cite{SVMs} implemented in LIB-SVM \cite{chang2011libsvm}  & Sampling interval, Acceleration along the x, y \& z & 15\% & 8\% \\ 
		
		\hline 
	\end{tabular} 
\end{table}

\subsection{Keystroke Based Authentication}

\begin{figure}[!h]
	\centering
	\includegraphics[width=90mm,scale=0.5]{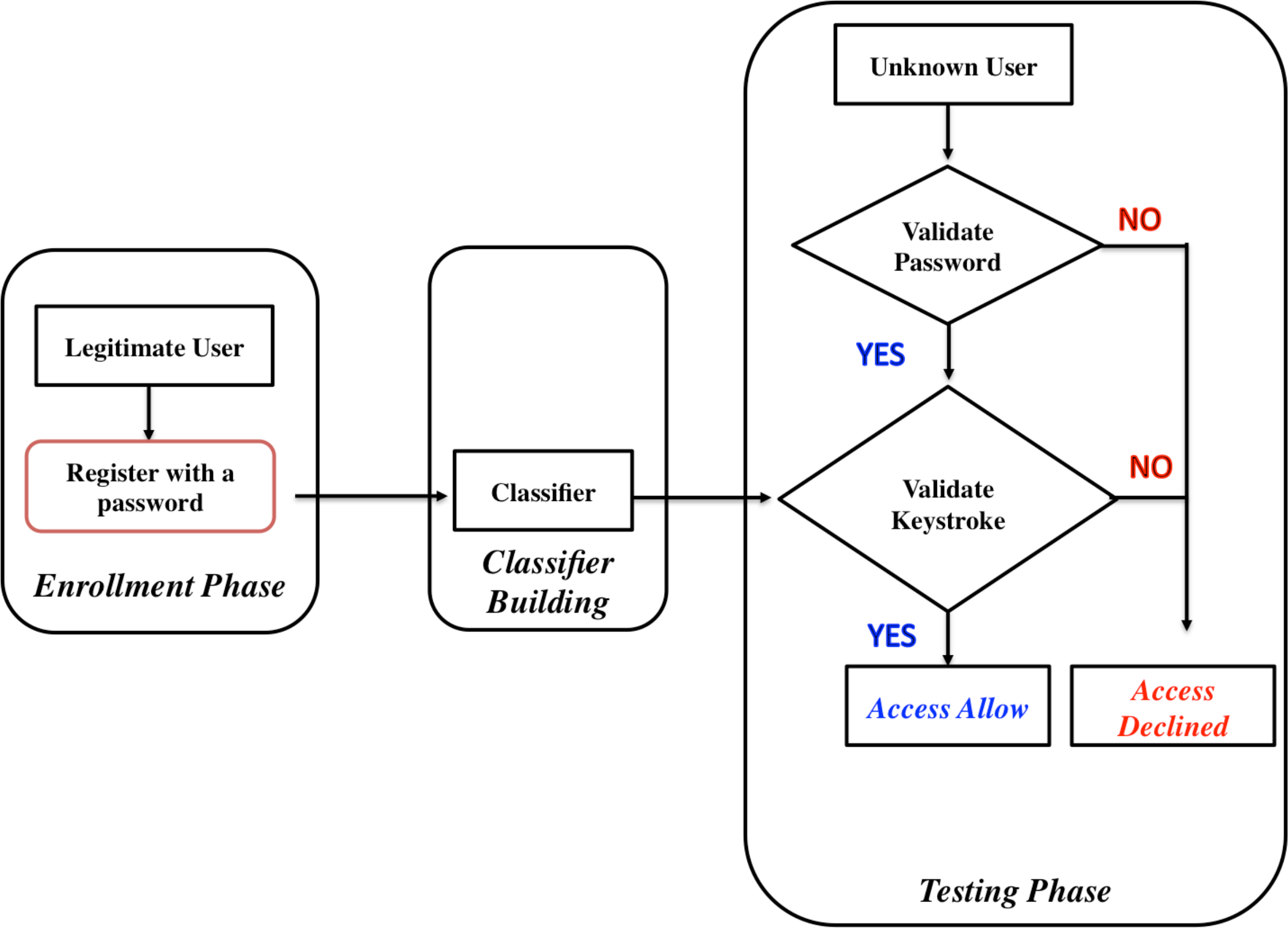}
	\caption{\textbf{General Architecture of Keystroke Authentication System}}
	\label{fig_KDA}
\end{figure}

Validating the nature of typing motion is one of the oldest methods to validate users. This technique analyzes keystrokes to determine authorized and unauthorized users. Typing motion or keystrokes can be used to detect and identify the user based on his/her manner of typing. Typing motion is divided into static and dynamic typing. In static typing, participants are asked to type a short and pre-defined text to compare motion information, while in dynamic typing, the subject is not required to type a specific string. He/she is free to type any text \cite{2010keystroke}. Figure. \ref{fig_KDA} shows the architecture of a keystroke dynamics-based user authentication system.

Table \ref{table_KF} outlines the main features used in keystroke based authentication of users.\\

\begin{table}[!h]
	\renewcommand{\arraystretch}{1.3}
	\caption{Keystroke Features}
	\label{table_KF}
	\centering
	\renewcommand{\arraystretch}{1.2}
	\begin{tabular}{|c|p{6cm}|}
		\hline
		\textbf{Feature} & \textbf{Description} \\

		\hline\hline
		Inter-time & The time between releasing a key and pressing of the next key. \\
		\hline
		
		Inter-key	& The interval between two successive keystrokes.\\
		\hline
		
		Hold-time & The time difference between pressing and releasing a single key.\\
		
		\hline
		
		Error-rate	& How many times backspace is pressed.\\
		\hline
		
		Distance	& The distance, represented in pixels, between two successively pressed virtual buttons.\\
		\hline
		Speed &	Computed from distance between two successively pressed virtual buttons divided by the sum of inter-times for each event.\\

		\hline 
	\end{tabular} 
\end{table}

\subsubsection{A mixture approach based on analysis of keystroke and handwriting}
Trojahn and Ortmeier\cite{trojahn2013toward} implemented a scheme that combines keystroke and handwriting analysis on smartphone and tablets. They performed two experiments to collect the data. In the first experiment, the authors asked 18 subjects to type a sentence containing two or more words ten times. They used the HTC Desire mobile phone with Android 2.2 (the dimension being 119 x 60 x 11.9 mm ) with a virtual keyboard. In the second experiment, the authors recruited 16 subjects to enter passwords eight times by using the HTC Desire HD with Android 2.3.5 (the dimension for the device is 123 x 68 x 11.8 mm). Trojahn and Ortmeier used WEKA \cite{WEKA} to evaluate their approach. They chose (J48) Decision Trees \cite{quinlan1996bagging}, Kstar \cite{cleary1995k}, Multi layer Perceptron (MLP) \cite{pal1992multilayer}, Radial Basis Function Network (RBFN) \cite{hwang1997efficient}, BayesNet \cite{friedman1997bayesian} and Naive Bayes \cite{rish2001empirical} as classifiers. \\

In the first experiment, the approach achieved EER of 2.0\%, while the second experiment obtained EER of 13.5\%. Decision Trees, Kstar, MLP, RBFN, BayesNet and Naives Bayes produce FAR of 2.03\%,3.49\%, 2.52\%,1.13\%,5.56\% and 19.29\%; respectively, and FRR of 2.67\%,2.21\%,3.0\%,4.47\%, 2.59\% and 1.8\%); respectively.\\

\subsubsection{Typing Authentication and Protection (TAP)}

Feng et al.\cite{feng2013continuous} implemented an approach using virtual key typing known as Typing Authentication and Protection (TAP). TAP divides an interaction into two phases. The login phase validates the identity of the user using password and biometric information. The application runs in the background to authenticate the user continuously. The second phase is the post-login phase, which is passive and works to validate the user when typing inputs.\\

The authors recruited 40 subjects (25 male and 15 female) and collect the data in a 3-week period. The authors distributed 10 Android devices to the subjects. In the first phase (login phase), the subjects entered passwords with 4 characters. 20 different passwords were used to test the performance of this phase. In the second phase (post-login), the subject typed sentences between 14 to 53 words. TAP extracted 40 features from key combinations, 41 features from key holding times without pressure and 41 pressure features. To evaluate TAP, the authors used three classifiers, which are Decision Trees \cite{quinlan1996bagging}, Random Forest \cite{pal2005random} and Bayes Net \cite{friedman1997bayesian}. In the first phase, the authors compared all classifiers in terms of time features, and time and pressure features, time and pressure features with touch. The performance for this phase is better when using Bayes Net, achieving FAR  10.4\%-16.9\% and FRR of 11.0\%. In the second phase, when the authors used input sequences of 60 characters, the Bayes Net algorithm has the best FRR of 0.27\%. However, Random Forest has the best FAR, which is 8.93\%.\\

\subsubsection{Keystroke inter keystroke latencies}

Buchoux and Clarke \cite{buchoux2008deployment} analyzed keystrokes on smartphones using a scheme with two stages, enrollment and authentication. The authors used two types of passwords, PIN and a complicated alphanumeric password. The main goal for this application is to capture key events and inter-keystroke latencies. The authors recruited 20 subjects for gathering data. Each subject had to type the password 20 times for the enrollment stage and 10 times for the authentication stage. The authors used two distance measures: Euclidean distance \cite{danielsson1980euclidean} and Mahalanobis distance\cite{de2000mahalanobis}, both with low processing requirements. They also used a Feed-Forward Multi-Layered Perceptron (FF-MLP) neural network \cite{bishop1995neural} to improve performance for their approach. Buchoux and Clarke evaluated their approach in terms of the inter-key feature, and found that the best result for the approach has FRR of 2.5\%.\\

\subsubsection{Using speed of finger movement}

Kambourakis et al.\cite{kambourakis2014introducing} introduced a method based on the speed and distance of finger movement on touchscreen devices. This method aims to improve analysis of traditional keystroke dynamics. The authors used four features as input data: hold-time, inter-time, distance and speed. To collect data, the authors recruited 20 subjects. Each subject has to repeat typing the password and other several phrases 12 times. The authors consider two scenarios. These scenarios are widely used in the literature on legacy keystroke analysis. For the first scenario, the subject has to type the password 7q56n5ll44, while in the second scenario, the subject has to type the phrase the quick brown fox jumped over the lazy ghost twelve times.\\

The authors selected three classifiers, which are Random Forest \cite{pal2005random}, k-NN \cite{knn} and MLP \cite{pal1992multilayer}. The authors found that the best results for the first scenario are achieved by using Random Forest \cite{pal2005random} with FRR of 39.4\%, FAR 12.5\% and EER 26.0\%, while in the second scenario the k-NN classifier achieves best results, which are FRR of 3.5\%, FAR 23.7\% and EER 13.6\%. The authors also evaluated their approach in terms of performance: time, CPU and memory. The CPU consumption is between 91\% and 100\%, and memory consumption fluctuates between 67.5\% and 80.2\% depending on the methodology, scenario and classification algorithm used. \\

\subsubsection{Keysens Approach: Micro-behavior modeling of soft keyboard interactions}

Draffin et al.\cite{draffinkeysens} introduced a new passive authentication approach called KeySens, based on micro-behavior of a user's interaction with his/her device's keyboard. The main goal is to illustrate how  some features can identify users. The authors analyzed the following features: how users touch keys at specific locations, the movement of fingers up and down, and the force of the touch and the area covered by finger. The authors recruited 13 subjects to participate in their experiment for three weeks and collect 86000 keypresses and 430000 touch data points. The authors trained a neural network classifier \cite{bishop1995neural} for authentication with the following features: touch-to-key offsets, sizes of finger touches, pressure, drag and hold time. KeySens is able to distinguish between legitimate and impostor users within 5 keypresses 67.7\% of the time with FAR of 32.3\% and FRR of only 4.6\%. After 15 keypresses, the KeySens is able to obtain 86\% FAR of 14\% and a FRR of only 2.2\%.\\  

\subsubsection{Keystroke dynamics-based user authentication using long and free text strings}

Kang and Choa \cite{kang2015keystroke} investigated keystroke dynamics-based user authentication (KDA) when long and free texts are typed using different input devices. This is different from most current approaches that focus on pre-defined and short inputs like passwords. Authentication using typing of a long piece of free text is affected by three factors, which are the type of input device, the length of the text strings and the type of authentication algorithm. To collect keystroke data, the authors used three input devices, a traditional PC keyboard, a soft keyboard and a touch keyboard. Each subject has to type a text with at least 3000 characters for each input device. The authors used one-class classifiers for their approach and build 12 one classifiers categorized into three groups. The classifiers used are the mean and variance equality test (MV test), Kolmogorov– Smirnov statistic (K–S statistic) \cite{lilliefors1967kolmogorov}, Cramér–von Mises criterion (C–M criterion) \cite{anderson1962distribution}, the distance between two digraph matrices (digraph distance; DD), Parzen window density estimator (Parzen) \cite{kwak2002input}, k-nearest neighbor detector (k-NN)\cite{knn}, and support vector data description (SVDD) \cite{tax2004support}. When the reference and test lengths are the shortest and equal to 100, the average EERs are the highest for all four experiments: 24:11\%; 30:61\%; 35:43\%, and 34:48\% for PC keyboard, soft keyboard, touch keyboard typed with one hand, and touch keyboard typed with two hands, respectively. The lowest EERs are reported with the longest reference and test lengths equal to 1000: 5:64\%; 14:10\%; 12:42\%, and 16:62\% for PC keyboard, soft keyboard, touch keyboard typed with one hand, and touch keyboard typed with two hands, respectively.\\

\subsubsection{Thumb-based keyboards for keystroke dynamic authentication}

Karatzouni and Clarke \cite{karatzouni2007keystroke} discussed the feasibility of keystroke analysis on thumb-based keyboards extracting two features when using two consecutive keys: inter-key and hold-time. The authors recruited 50 subjects to participate in this approach. The subjects were asked to enter 30 messages. Each message is designed to ensure certain requirements. The authors used a Feed Forward Multilayer Perceptron Neural Network (FF-MLP) \cite{bishop1995neural} as classifier. The proposed scheme with inter-keystroke times obtains an average EER of 12.2\%. However, the hold-time characteristic is not very useful in the analysis of keystroke interface because it has a higher EER, which is between 36.8\% and 50.02\%.\\

\subsubsection{Identify users based on digraph features}

Zahid et al.\cite{zahid2009keystroke} discussed the keystroke dynamics of the users and use 6 different keystroke features to identify users. The features are Key hold time, Digraph time, Horizontal Digraph, Vertical Digraph, and Non-adjacent Horizontal and Vertical Digraph. The authors recruited 25 subjects for this study for 7 days. The authors used the following classifiers: Naive Bayes \cite{rish2001empirical}, Back Propagation Neural Network (BPNN) \cite{paola1995detailed}, Radial Basis Function Network (RBFN) \cite{hwang1997efficient}, Kstar \cite{cleary1995k} and J48 Decision Tree \cite{quinlan1996bagging}. The authors used the implementations of these algorithms in WEKA \cite{WEKA}. Most classifiers produce an FAR of 30-40\%, and FRR of most of the classifiers is approximately 30\% or higher. These results are not acceptable. Therefore, the authors used a fuzzy classifier with much better FAR and FRR of approximately 18.6\% and 19.0\%, respectively. However, the accuracy was still low, and as a results, the authors incorporated a PSO and GA-based optimizer into the fuzzy classifier \cite{kuncheva2000fuzzy}, and achieve EER lower than 2\% after detection mode and FRR close to 0\% after verification of PIN.\\ 

\subsubsection{Using sensors with a virtual software keyboards to identify users}

Gascon et. al \cite{gasconcontinuous} developed a software keyboard application for the Android OS to collect behavioral biometrics. They implement a keyboard with three sensors, which are accelerometer, gyroscope and an orientation sensor. Each one of these sensors measure in three dimensions: X, Y, and Z. They recruited 315 participants to write pre-defined short texts. The participants categorized into two groups. The first group consisted of 303 participants who type the text one time only. They represented impersonators who can possibly steal devices if these devices are found unlocked. The second group consists of 12 authorized users, who were asked to type the text 10 times so that enough typing motions can be obtained. \\

To ensure that only motion data related to typing behavior is used to analyze a user's profile, Gascon et al. processed the sensor signals according to certain time constraints. Users were observed for $T$ seconds while typing. For example, a data point is recorded every $T$ seconds while a user is typing the text. If the user does not type any letter in $T_{stop}$ seconds, it assumes that the user is finished typing the text. Once the user decides to retype text, the new value of $T$ seconds is added to the previous $T$. This process collects all motions that are related to typing.\\

Gascon et al. used Linear Support Vector Machines \cite{SVMs} to identify the behavioral fingerprint of each user to distinguish between authorized users and impersonators. The average classification performance of classifiers for non-identifiable users has a considerably high FPR of 35\% while reaching only a TPR of 58\%. The value of FPR is too high. This causes the device to be locked continuously, which means the user has to validate each time he/she wants to use the device. This action is incompatible with the purposes of continuous authentication. Therefore, the FPR must be reduced to avoid this issue. On the other hand, with regards to clearly identifiable users, the average classification performance achieves a TPR of 92\%, and FPR of only 1\%. \\

Table \ref{table_Key} summarizes current studies that use typing motion approaches.\\

\begin{table*}[h]
	\renewcommand{\arraystretch}{1.3}
	\caption{Examples of use of Keystroke Biometrics}
	\label{table_Key}
	\centering
	\renewcommand{\arraystretch}{1.2}
	\begin{tabular}{|p{1.7cm}|p{1cm}|p{2.1cm}|p{2.7cm}|p{1.6cm}|p{1.7cm}|p{1.5cm}|p{1.5cm}|p{1.5cm}|}
		\hline
		\textbf{Study} & \textbf{Dataset} & \textbf{Classifiers/Metrics} & \textbf{Used Feature} & \multicolumn{5}{c|}{\textbf{RESULT}}\\
		
		\cline{5-9}
		&&&& \textbf{EER} & \textbf{FAR} & \textbf{FRR}  &  \textbf{CPU consumption} & \textbf{Memory consumption}\\
		\hline\hline
		
		Trojahn and Ortmeier\cite{trojahn2013toward},2013 & 16 &  J48 \cite{quinlan1996bagging}, Kstar \cite{cleary1995k}, MLP \cite{pal1992multilayer}, RBFN \cite{hwang1997efficient}, Bayes Net\cite{friedman1997bayesian} and Naive Bayes \cite{rish2001empirical} & Diagraph, Pressure, Finger size& 
		2.0\% (1st exp) 13.5\% (2nd exp) & J48 2.03\% & J48 2.67\% & - & -\\      
		
		\hline
		
		Feng et al.\cite{feng2013continuous}, 2013 & 40 & J48 \cite{quinlan1996bagging}, Random Forest \cite{pal2005random} and Bayes Net \cite{friedman1997bayesian} &Holding time, Tactile pressure, Virtual key combinations   & - & Bayes Net 10.4\% - 16.9\% (1st phase) Random Forest 8.93\% (2nd phase) , 0.27\% (2nd phase) &  - & - & -\\
		
		\hline 
		
		Buchoux and Clarke \cite{buchoux2008deployment}, 2008 & 20 & Euclidean distance \cite{danielsson1980euclidean}, Mahalanobis distance \cite{de2000mahalanobis}, FF-MLP   \cite{bishop1995neural} & - & - & - & 2.5\% &  - & -\\
		\hline

	 Kambourakis et al.\cite{kambourakis2014introducing}, 2014 & 20 & Random Forest \cite{pal2005random}, k-NN \cite{knn} and MLP \cite{pal1992multilayer} & Hold-time, Inter-time, Distance, Speed & Random 26.0\% (1st sce), 
		k-NN 13.6\% (2nd sce) & - & - & 91\%-100\% & Fluctuates between 67.5\% -80.2\%    \\
		\hline 
		
		Draffin et al.\cite{draffinkeysens}, 2014 & 13 & Neural network classifier \cite{bishop1995neural} & Location pressed on key, Length of press, Pressure, Size of touched area, Drift & - &  14\% & 2.2\% &  - & -\\
		
		\hline
		
		Kang and Choa \cite{kang2015keystroke},2015 & 35 & 12 one-class learning &Length of the text, Start\& end key, Time between two key presses & PC keyboard 5:64\% (length text 1000)  & - & - & - & - \\

		\hline

		Karatzouni and Clarke \cite{karatzouni2007keystroke},2007 & 50 & FF-MLP \cite{bishop1995neural} &Inter-keystroke latency, Hold-time& Inter-keystroke12.2\%, Hold-time between 36.8\% -50.02\% & - &- &- & - \\
		
		\hline 
		
		Zahid et al.\cite{zahid2009keystroke}, 2009 & 25  & Naive Bayes \cite{rish2001empirical}, BPNN \cite{paola1995detailed},  RBFN \cite{hwang1997efficient}, Kstar \cite{cleary1995k}  and J48  \cite{quinlan1996bagging} & Hold time, Times backspace is pressed, Horizontal \& Vertical Digraphs, Non-Adjacent Horizontal \& Vertical Digraphs & Lower than 2\% in detection mode & Close to 0\% after verification of PIN & - & - & - \\
		
		\hline 
		
		Gascon et. al \cite{gasconcontinuous}, 2014 & 315 & Linear Support Vector Machines \cite{SVMs} &Simple Statistics, Spline Coefficients, Spline Simple Statistics, iFFT Spline Features, iFFT Signal Features & - & - & Non-identifiable 35\%, identifiable users 1\% & Non-identifiable 58\%, identifiable users 92\%  & - \\  
		
		\hline
		
	\end{tabular}
	
\end{table*}

\subsection{Touchscreen Based Authentication}

Smartphone technology has undergone rapid evolution in the recent past. One main innovation is the touchscreen, which has become the essential input medium. A touchscreen is an electronic visual display  for inputs and outputs. In 2013, global touchscreen shipments reached 1.75 billion pieces of, which approximately 1.28 billion (73\%) are for handsets, which is a 14.2\% on-year increase\cite{TJY}. Analysis of the nature of touchscreen usage has become one of the most interesting areas to validate and distinguish among users. In this section, we discuss several studies that aim to enhance security and defeat possible attacks.\\    

\subsubsection{Password Patterns} 

De Luca et al.\cite{de2012touch} introduced an approach to improve the level of security using password patterns with an implicit authentication layer. They performed two studies, a short-term lab study and a long-term study using higher granularity. In the first study, they explored simple security procedures (e.g., using a horizontal stroke), while in the second study, they used password patterns, each enhanced with passive authentication. These studies used Android applications to gather information from participant inputs on touchscreens. De Luca et.al.'s application records all data available from a touchscreen device: pressure (how hard the finger presses), size (area of the finger touching the screen), X and Y coordinates, and time of contact. Evaluation was based on 26 participants, with 645 valid authentication attempts and 2790 attacks. The authors used Dynamic Time Warping (DTW)\cite{DTW} for analysis. This study produced 398 True positives, 231 False positives, 852 True negatives, 92 False negatives with False Rejection Rate of 19\%, False Acceptance Rate of 21\% and an average accuracy equals to 77\%.\\ 

\subsubsection{Non-Intrusive Approach via Tapping Behavior}

\begin{figure}[!h]
	\centering
	\includegraphics[width=3.5in]{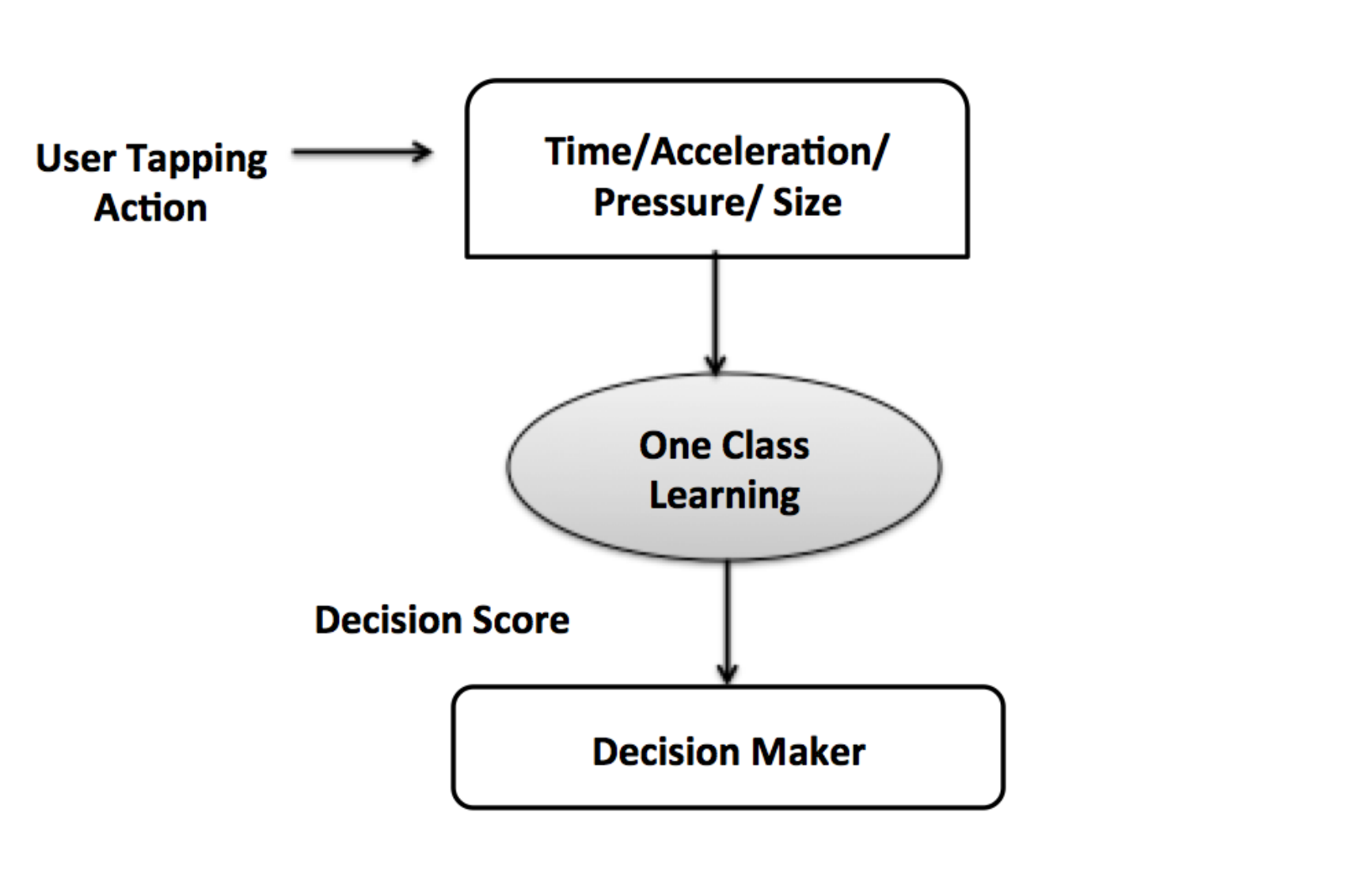}
	\caption{Non-Intrusive User Verification Architecture}
	\label{Fig_zheng}
\end{figure}

Zheng et al.\cite{zheng2012you} proposed a mechanism called non-intrusive user verification. The authors studied behavioral authentication by analyzing the way a user touches the phone, taking into account different sensors such as accelerometer, gyroscope and touch screen sensor. Fig. \ref{Fig_zheng} outlines the architecture of the system. 

The authors recruited 80 subjects to participate in this study. They used sensors (accelerometer, gyroscope and touchscreen sensors) to record a user's behavior. For example, information such as acceleration, pressure, size of touch area and passage of time were collected using empirical data on both 4-digit and 8-digit PINs by employing tap behaviors to verify passcodes of participants. The authors used five different PINs for testing: 3-2-4-4, 1-1-1-1, 5-5-5-5, 1-2-5-9-7-3-8-4 and 1-2-5-9-8-4-1-6. The participant has to enter the PIN error-free, and the PIN has to be entered at least 25 times. This gathered enough error-free action for analysis. The authors used one-class learning based on the notion of nearest neighbor distance \cite{1C}. If the distance is large, the probability is that the person holding the device is an imposter. The proposed approach obtained EERs for all five inputs, which were between 3.58\% for the PIN 3-2-4-4 and 7.34\% for 5-5-5-5. The other results were 6.96\% for 1-1-1-1, 4.55\% for 1-2-5-9-7-3-8-4 and 4.45\% for 1-2-5-9-8-4-1-6.

\subsubsection{Graphic Touch Gesture Feature}
\begin{figure}[!h]
	\centering
	\includegraphics[width=9cm,height=9cm,keepaspectratio]{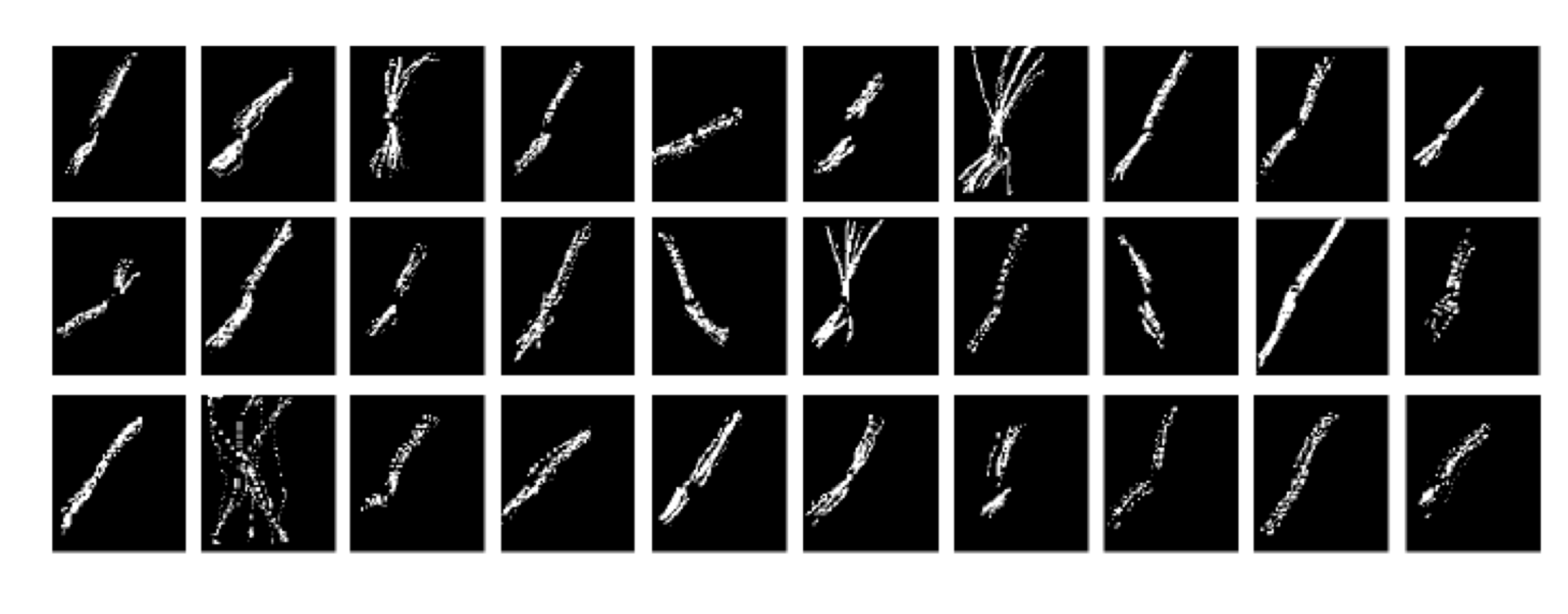}
	\caption{\textbf{Over 300 touch traces of zoom-in gestures from 30 users. From \cite{zhao2013continuous}, With permission.}}
	\label{Fig_trace}
\end{figure}

\begin{figure}[!h]
	\centering
	\includegraphics[width=9cm,height=9cm,keepaspectratio]{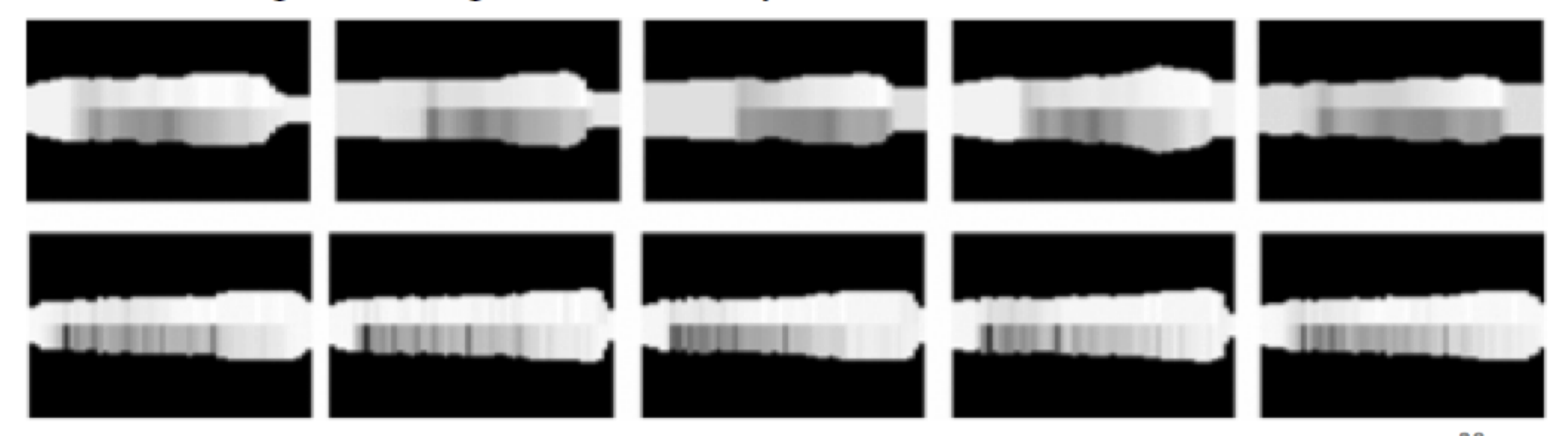}
	\caption{\textbf{Examples of GTGF extraction. From  \cite{zhao2013continuous}, With permission.}}
	\label{Fig_GTGFE}
\end{figure}

Zhao et al.\cite{zhao2013continuous} implemented an approach called the Graphic Touch Gesture Feature (GTGF). GTGF is implemented on the Android OS. This application captures and observes touch gestures using a standard API of the Android system. When users start touching the screen, the application begins to record the trace as raw touch samples from the API. In each trace, the system can monitor an event flag, absolute event time in milliseconds and (x,y) coordinates. It can also capture tactile pressure for each contacted finger. The authors used 6 sessions, which were divided into two groups. The first group, containing only one session, called UH-TOUCHv1, evaluated features and verified the design of the authentication system. The second group called UH-TOUCHv2, included 5 sessions and used continuous authentication. The authors asked 30 subjects to participate in this experiment. The authors collected more than 300 touch gestures from subjects. Figures . \ref{Fig_trace} and  \ref{Fig_GTGFE} outline examples of extracted features. Each sub-figure contains 10 traces from a single user.\\ 

\begin{figure}[!htp]
	\centering
	\includegraphics[width=7cm,height=7cm]{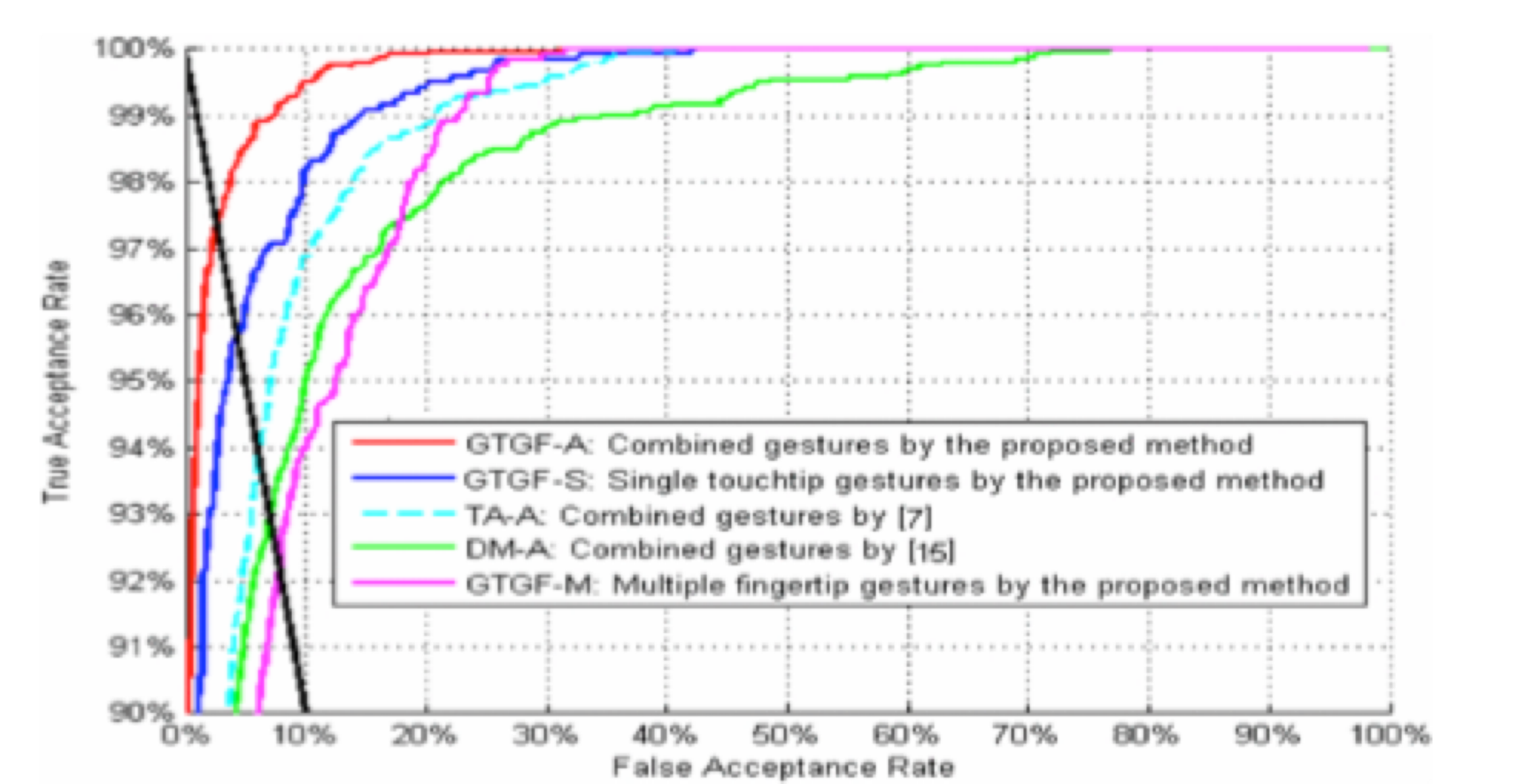}
	\caption{\textbf{A comparison on ROC curves for different fusion schemes. From \cite{zhao2013continuous}, With permission.}}
	\label{Fig_curv_touch}
\end{figure}

Zhao et al. evaluated the performance of GTGF with the 3 metrics including $L_1$ distance \cite{huang1997image}, $L_2$ distance or Euclidean Distance \cite{danielsson1980euclidean} and normalized cross correlation \cite{lewis1995fast}. The authors found the following values for the metrics: 11.28\%, 12.38\%, and 17.29\% for $L_1$ distance, $L_2$ distance and the normalized cross correlation, respectively. This means that $L_1$ distance works better than the other metrics. The authors used the UH-TOUCHv2 dataset for assessing the approach, evaluating it by using ROC curves for four fusion schemes, which are: GTGF--A: a scheme that computes all six gestures from the proposed approach, GTGF--S: a scheme that used only single gestures (slide up, down, left and right), GTGF--M: a scheme that used multiple gestures (Zoom in/Zoom out), TA-A: a scheme that used six gestures from in the approach in \cite{frank2013touchalytics} and DM-A: which used six gestures from \cite{sae2012investigating}. Fig. \ref{Fig_curv_touch} shows a comparison using ROC curves of all six fusion schemes. The best performance was achieved by the GTGF-A approach (which used EER) at 2.62\% while the worst performance was from GTGF--M approach at 7.81\%. The results for the other three approaches are: GTGF--S 4.31\%, TA-A 6.07\% and DM-A 7.06\%. The authors justified the good results due to the use of tactile pressure, which contains extra clues about the subject's muscle behavior and the extraction of dynamic movement and dynamic pressure from the touch gesture during feature extraction. \\

\subsubsection{Re-Authentication Model Based on Movements of a User's Finger}
\begin{figure}[!h]
	\centering
	\includegraphics[width=3.5in]{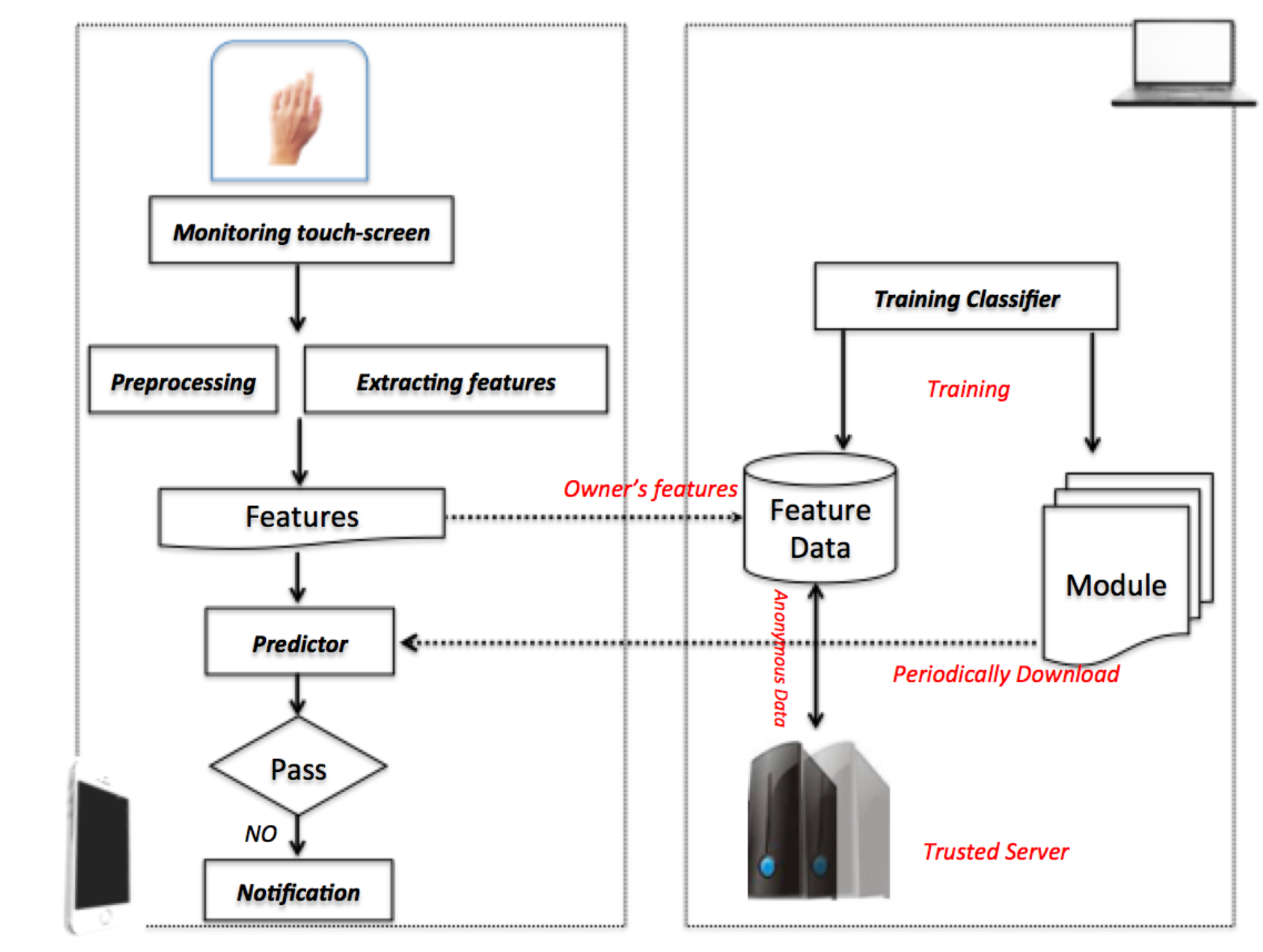}
	\caption{\textbf{Architecture of Re-authentication Model}}
	\label{Fig_arch}
\end{figure}

Li et al. \cite{li2013unobservable} proposed a method for smartphones to re-authenticate users. This method focuses on examining the movements of a user's fingers and learning  movement patterns. This process keeps running in the background and continuously monitors the current user's finger movements, in order to compare these movement patterns to the owner's patterns. This approach does not need assistance from the user to re-authenticate behavior. The methodology can be adapted to other platforms, and it clearly shows that gathering information on finger movement to construct a continuous re-authentication model, without the user's knowledge is practical. Figure. \ref{Fig_arch} shows the architecture of the re-authentication model.\\

Li et al.'s \cite{li2013unobservable} model, used two Motorola Droid phones to gather data. The phones have the following features: 550MHz A8 processor, 256MB memory, 16GB SD-CARD and Android 2.2.2 OS. To collect gesture data, 75 participants used the smartphones without constraints for several days. They were allowed to access websites such as online social networks and were able to download OSNs' applications to these devices.\\

Users had a choice of using or ignoring security mechanisms. The participants were categorized into two groups. The first group of 28 was asked to use the smartphones with at least 150 sliding up, 150 sliding down, 150 sliding right, 150 sliding left, and 300 flip gestures. The second group of participants included 47 users who use the phones for a total of fifteen minutes. \\

The authors used an open-source Android application that observes movements of the user's fingers and categorizes them into: sliding up or down, sliding right or left and flipping. The monitoring is done in the background. This application also records touch events, reading the data sent to the Linux kernel, which is responsible for forwarding data to the upper layer of the Android library. Since this data can be read only from lower layers, the program needs root privilege.\\

The authors used SVM \cite{SVMs} for training purposes. Li et al. \cite{li2013unobservable} computed the average time interval for each gesture type in the collected user data. For example, a sliding up gesture occurs every 8.24 seconds in portrait mode. This means the approach has the ability to re-validate authenticity of a user by considering sliding up gestures every two minutes. \\ 

The authors evaluated overhead, accuracy and performance for their approach and found that the results of the experiments were affected by many factors. First, to assess re-authentication frequency, they kept track of the number of gestures needed to perform re-authentication. The authors found that FAR and FRR were stable once the block-size equaled 14. They also observed that, for the tap gesture, FAR and FRR become constant once the block size was approximately 20. The tap gesture exhibits the worst performance with the highest FAR and FRR among the gestures. The second factor that affects the accuracy of module classification is the size of the training data set.
Results from performance testing of the classification module exhibited a behavior where the accuracy initially rose with the size of the training set, but it then reached a maximum and decreased thereafter. The researchers recorded 500 as the optimum point for modules with a single gesture and 300 for the category of modules with combination gestures. Another thing to note is that the authors found that modules with tap as a supplementary gesture produced better results than those with only sliding gestures. Modules with sliding left gestures were deemed to have the best results, while those with portrait and landscape sliding down gestures had the worst results. The results obtained were FAR of 4\% and FRR of 4\%, which were the best results the authors obtained.\\

\subsubsection{Touchalytic system continuous authentication}

Frank et al.\cite{frank2013touchalytics} discussed the question of whether or not touchscreen input is effective as a behavioral biometric for continuous authentication. They demonstrated that simple touch movements are enough to verify a user. They introduced an approach, which is known as the Touchalytic system for data collection. Frank et al.'s \cite{frank2013touchalytics} system gathered touch data from Android phones, by creating an application for reading documents and comparing multiple photos and images. To collect data, users were asked to read a few documents from Wikipedia. The users were also asked to spot differences among photos presented to them. While users read or browsed images, their actions were observed to extract raw movement features. The raw features can be event codes, which include finger up, down, move or multi-touch events. Data collected consists of the pressure on the screen, the area of the screen covered by the finger, the finger orientation with respect to the screen and the screen orientation. For the primary study, the overall experiment time ranged between 25 and 50 minutes per subject. A single reading trial usually takes between ten and fifteen minutes while each image comparison trial takes approximately 3 to 4 minutes. There were 41 participants in the study and 4 different smartphones with similar specifications are used. The authors extracted 30 touch features once the user logged into the devices. They used k-NN \cite{knn} and Gaussian kernel Support Vector Machine \cite{rasmussen2006gaussian} for training users profiles. Using Touchalytic, the authors obtained EER between 2\%-3\% for intersession authentication, EER 0\% for intra-session, and under 4\% for tests carried out one week after the enrollment phase. \\

\subsubsection{Identity Protection Service (TIPS)}

\begin{figure}[!h]
	\centering
	\includegraphics[width=4in]{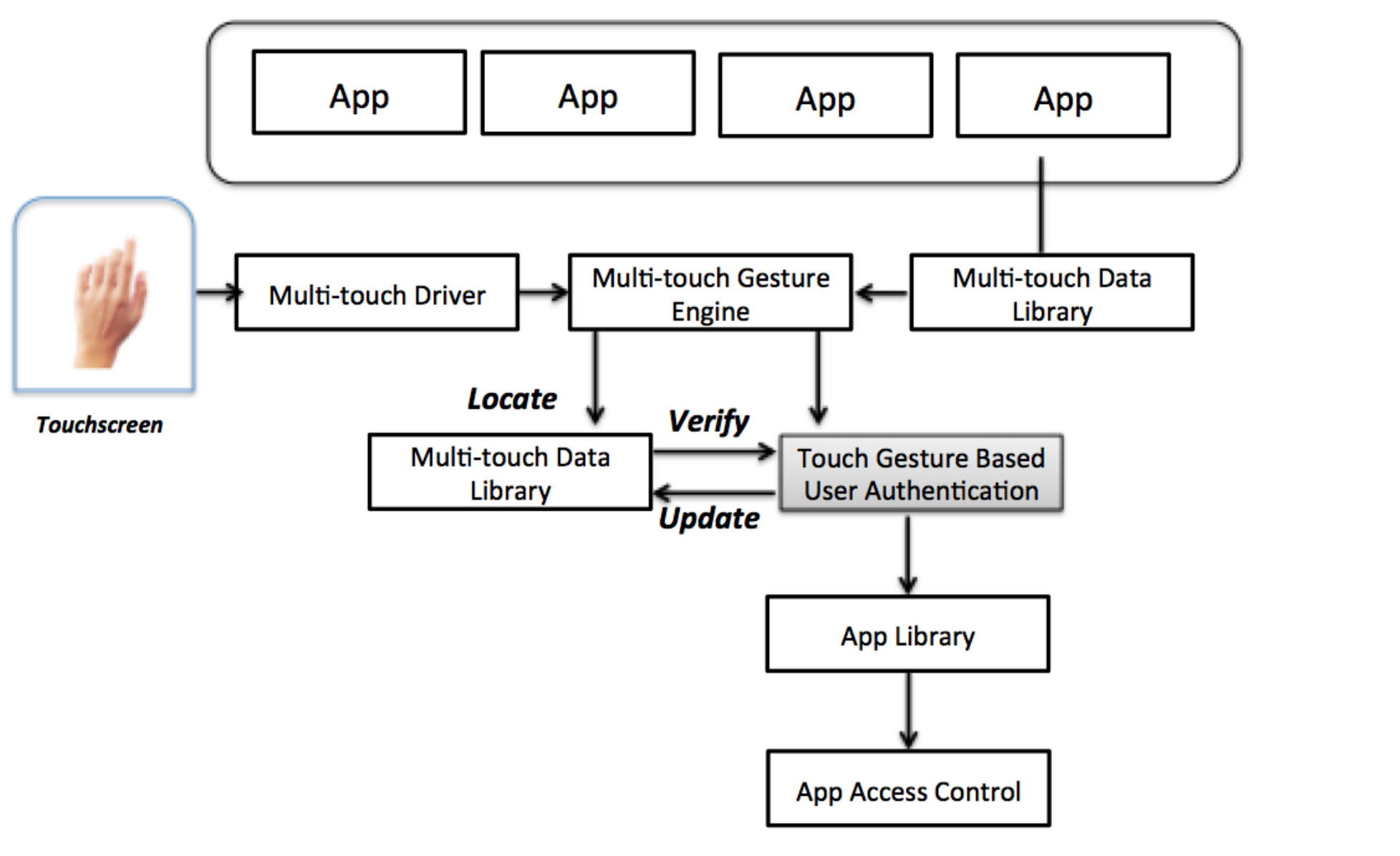}
	\caption{Structure of TIPS Scheme}
	\label{Fig_DTIPS}
\end{figure}

Feng et al.\cite{feng2014tips} presented a mechanism for touch analysis called Identity Protection Service (TIPS). The objective of this approach is to resolve three main issues: dealing with data in an uncontrolled environment, enhancing accuracy and achieving real time recognition. TIPS collects behavioral and contextual features and works by gathering gesture input data. This approach runs two contextual applications, which are a multi-touch driver and a context listener. TIPS always runs in the background and checks touch data continuously. It incorporates information about contextual applications to perform user authentication. Figure. \ref{Fig_DTIPS} shows the architecture of the TIPS strategy.

For testing, the authors allocated 23 smart devices (8 Galaxy S3, 3 Galaxy S4, 12 Nexus 4) to 13 users who used 100 gestures. The experiment consisted of two phases. First, in the off-device simulation phase, Matlab\footnote{http://www.mathworks.com/products/matlab/} was used to construct and analyze off-device touchscreen data for 3 weeks. The second phase was an on-device testing phase, in which online training and testing modules were combined into the TIPS service. 2,000 touch gestures were collected for each user over a week. The user has the ability to set the mode to either training or notification, and also the authentication length parameter. In order to classify a variety of touchscreen gestures, the authors used a combination of the One Nearest Neighbor (1NN) classifier \cite{garcia2012prototype} and Dynamic Time Warping (DTW) \cite{DTW}.\\

Feng et al.\cite{feng2014tips} assessed their approach in two situations. First, they evaluated the off-device situation. They estimated performance by setting various authentication lengths. They built four different template libraries with different sizes, and found that once they minimize the size of the templates, accuracy decreases. They also observed that the accuracy improves once the authentication length is increased. The approach reached 91\% for TP and 93\% for TN. The authors also calculated the power consumption of their approach, and the average of the energy usage was 88mW. The battery usage was below 6.2\%.\\

\subsubsection{Authenticating using Keystroke Dynamics and Finger Pressure}

Saevanee and Bhattarakosol\cite{saevanee2009authenticating} implemented a method, which tries to detect and recognize users in terms of hold-time, inter-key and finger pressure. This approach combined the use of both keystroke and touchscreen patterns. The authors asked 10 subjects to type their phone number repeatedly 30 times. Values like finger pressure and finger position were recorded every 20 ms. They collected 3,000 values of hold-time, 2,700 values of inter-key and 3,000 values of finger pressure. For classification, the authors used Probabilistic Neural Networks (PNN)\cite{PNN90}, which estimate the probability density functions from a set of training patterns. The authors evaluated this approach in two ways. First, they combined three metrics producing an EER of 9\%. Second, when hold-time and inter-key metrics were used together, the EER was 29\%. These results show that the values of EER are not good enough because this approach rejects legitimate users with high probability, leading to constant re-authentication of the user.\\

\subsubsection{Progressive Authentication when Entering PINs with face and speaker Recognition}

Riva et al.\cite{riva2012progressive} measured the user's frequency of entering a PIN with or without progressive authentication. The aim of this work is to reduce the authentication overhead on smartphone devices, and to make authentication a viable solution for users who currently do not use security locks on their devices. They also considered the number of unauthorized authentications. 
To validate the system, the authors recruited 20 subjects. The authors compared SVM \cite{SVMs}, Decision Tree \cite{quinlan1996bagging} and Linear Regression \cite{seber2012linear} and computed Precision and Recall, and found that SVM \cite{SVMs} achieved better results than the other two algorithms. The approach decreased the number of times a user is requested to enter a PIN by 42\%. The recall for SVM was 92.5\%. SVMs \cite{SVMs} produced false authentications at a rate of less than 8\%.\\ 

The approach enhances performance from both a convenience and a security perspective by using a high/low risk factor $F$. Risk factor $F$ is a variable that is associated with either increased or decreased risk. Table \ref{table_FAs} shows that the authors used different models with different risk factors $F$. It compares risk factors in terms of FA and FR, between private and confidential applications versus when risk factor is equal to 1 (represents baseline without loss function and shows the inference model for security).    

\begin{table}[H]
	\renewcommand{\arraystretch}{1.3}
	\caption{Comparison among risk factors in terms of percentage of false authentications and false rejections for private (FA Priv and FR Priv) and confidential (FA Conf and FR Conf) applications}
	\label{table_FAs}
	\centering
	\begin {tabular}{|c|c|c|c|c|}
	\hline
	\bfseries Risk Factor & \bfseries FA pv & \bfseries FR pr & \bfseries FA Con & \bfseries FR Con \\
	\hline \hline 
	0.05 & 3.3 & 0.0 &  57.7 & 100.0 \\
	\hline 
	0.2 & 3.6 & 0.0 & 55.8 & 100.0 \\
	\hline 
	1 & 4.9 & 0.0 & 53.5 & 98.4 \\
	\hline 
	5 & 5.8 & 0.0 & 39.9 & 96.8 \\
	\hline 
	20 & 16.1 & 0.0 & 34.4 & 96.8\\
	\hline
\end{tabular}
\end{table}
\lipsum[0-0]

Table \ref{table_FAs} shows that increases in $F$ can lead to rise in the percentage of FAs for private applications from 4.9\% to 16.1\%. In contrast, when $F$ = 20, FR comes down to 34.4\%. In such a case, the user has to enter the PIN for private applications one out of three times, while he or she must enter the PIN for confidential applications every time. In this situation, the FR confidential is 96.8\%.\\

The authors calculated accuracy in different settings, such as using and not using sensors, using and not using face and speaker recognition. The model's accuracy varied from 83\% to 100\% when all sensors were used and from 76\% to 99\% if only accelerometers were used. Without an added Gadgeteer\footnote{http://www.netmf.com/gadgeteer/} sensor, a platform used to construct small electronic devices, the model still provides for high accuracy. For face recognition, a 94\% accuracy was achieved with minimal discrepancy among users. A small decreased in accuracy is observed in fluctuating light situations. Varying results were obtained in the accuracy of the speaker identification model; the average accuracy was around 59.3\% for 3 users, with a high of 83.7\%, and a low of 0.4\%. The varying results can be attributed to different conditions during the time of recording (e.g., distance between the user and speaker for example; if a user is talking or the device is in the pocket), and background noise such as noise from air conditioner and keyboard typing.\\

\subsubsection{Authentication based on Curve Training}

Burgbacher et al.\cite{burgbacher2014behavioral} introduced a method that gathers information from users as they trace curves. First, the approach randomly generated 19 curves where the starting and ending points were shown to indicate the trace direction. Then, each user was asked to trace this curve with his/her finger. The authors extracted X-Y coordinates to represent finger location and timestamps for each location. The authors recruited 42 participants, 12 female, 30 male. 6 of the subjects used their left hand whereas the rest used right hand. To eliminate any sensor noise and outliers, they divided the curves into small segments in a pre-processing step. Burgbacher et al. \cite{burgbacher2014behavioral} extracted curve segments using k-NN\cite{knn} and compute the similarity between two segments by employing Dynamic Time Warping (DTW) \cite{DTW}. To evaluate this approach, the authors performed a 10-fold cross-validation. They obtained EER between 10\% -20\%.  \\

\subsubsection{Fingergestures Authentication System using Touchscreen (FAST)}

Feng et al.\cite{feng2012continuous} introduced a touchscreen-based approach, known as Fingergestures Authentication System using Touchscreen (FAST). The authors used digital gloves that contain a sensor (IMU digital combo boards ITG3200/ADXL345), with 6 degrees of freedom to collect the movement of fingers. The system collects data when a user flicks, spreads, taps, types or enters phone numbers and draws shapes. FAST gathers X-Y coordinates, speed, pressure and directions of finger motions.\\

To evaluate FAST, Feng et al.\cite{feng2012continuous} recruited 40 participants. They divided subjects into two groups. The first group of 11 subjects used sensor gloves while the second group did not wear anything. Both groups performed six gestures: down to up swipe, up to down swipe, left to right swipe, right to left swipe, zoom-in, and zoom-out. The authors collected 53 features for each touch gesture. They also captured 36 triaxial angular rate features.\\

The authors used three classification algorithms for evaluating  FAST: (J48) Decision tree \cite{quinlan1996bagging}, Random Forest \cite{pal2005random} and Bayes Net \cite{friedman1997bayesian}. The authors compared the three classifiers in two situations, which were with and without gloves. They compared all three algorithms in terms of FAR and FRR. The Random Forest Classifier has the best result in terms of FAR while the Bayes Net classifier has the best results in terms of FRR. When using the Bayes Net classifier for single touch gestures, FAST obtained FAR of 11.96\% and a FRR of 8.53\% when not using external sensors, while it got FAR of 2.15\% and a FRR of 1.63\% when using sensor gloves. For multi-touch gestures, the authors also applied these classifiers to the system. FAST with Bayes Net classifier obtained FAR of 14.02\% and FRR of 18.92\%. This result shows that FAST does not achieve the requirement of a good authentication design. As a result, the authors implemented the system based on a sequence-based authentication mechanism which uses time threshold. This new approach was achieved a FAR of 4.66\% and a FRR of 0.13\%. \\

\subsubsection{LatentGesture}

Saravanan et al.\cite{saravanan2014latentgesture} introduced an approach called LatentGesture that analyzes events for touch interactions with user interface elements like buttons, checkboxes and sliders. It extracts features such as position on screen, time, frequency and pressure. To collect data, Saravanan et al. recruited 20 subjects. All subjects used two devices, a Nexus 4 phone and Nexus 7 tablet. The authors use Lib-SVM \cite{chang2011libsvm} and Random Forests \cite{pal2005random} to perform unary classifications. The authors evaluated the approach in two situations, multi-and single-users authentications. For single-user authentication, the accuracy achieved was 97.9\% on the phone and 96.79\% on the tablet. For the multi-user authentication, the accuracy achieved was 100\% on the tablet and 97.78\% on the phone. \\

\subsubsection{Authentication using only 3 Gesture Features}

Shahzad et al. \cite{shahzad2013secure} implemented a Gesture based Authentication scheme for secure unlocking of Touchscreen devices (GEAT). GEAT uses several sensors to determine how users input data. These sensors are finger velocity, device acceleration and stroke time. GEAT has three phases. The first phase collects gestures from the touchscreen. In this phase, the subject is asked to touch the screen between 15-25 times. The second phase is based on extracting features from the data collected. The third phase constructs models to classify each gesture using machine learning.\\

The authors used Samsung smartphones and recruited 50 subjects. Each subject used the device for approximately 7-10 days. 15009 gestures were gathered during a period of 5 months. The authors employed Support Vector Distribution Estimation (SVDE) \cite{keerthi2003asymptotic} with the Radial Basis Function (RBF) kernel \cite{hwang1997efficient} for classification. They used the open source implementation of SVDE in libSVM\cite{chang2011libsvm}. The authors used the EER metric for evaluation. The results show that by using three gestures and 25 training samples, GEAT can achieve an average Equal Error Rate (EER) of 0.5\%.  \\

\subsubsection{Touch Dynamics Authentication Scheme}

Meng et al.\cite{meng2013touch} implemented a scheme that uses 21 features for authentication, called the touch-dynamics-based authentication system. It has three phases, which are data collection or acquisition of data once a user touches this screen and conversion of the data into meaningful features; behavior modeling to analyze the data, e.g., extracted features to produce signatures for legitimate users; and behavior comparison to compare between current behavior and produced signatures to make a decision if the user is legitimate or not.\\ 

The authors used Google/HTC Nexus One Android phone and recruited 20 subjects. Data collected include X-Y coordinates, timestamps, and input characteristics like single or multi-touch. Each subject finished 6 sessions within 3 days. In each session, the subject spent at least 10 minutes. The authors gathered 120 sessions for all subjects.\\ 

Meng et al.\cite{meng2013touch} used five classification algorithms to validate the scheme. These classifiers were: (J48) Decision Tree \cite{quinlan1996bagging}, Naive Bayes \cite{rish2001empirical}, Kstar \cite{cleary1995k}, Radial Basis Function Network (RBFN) \cite{hwang1997efficient} and Back Propagation Neural Network (BPNN) \cite{paola1995detailed}. The authors used WEKA\cite{WEKA} implementations of the classifiers and evaluated the system in terms of EER. The best results were from RBFN \cite{hwang1997efficient} and BPNN \cite{paola1995detailed}, which were 7.71\% and 11.58\%, respectively although the results are still very high. The other classifiers had ERR between 15\% and 24\%. To enhance the results, the authors implemented a hybrid classifier called PSO-RBFN, which decreased the error rate to 2.92\%. \\

\subsubsection{Continuous and Passive Authentication via Touch Biometrics}     

Xu et al.\cite{xu2014towards} discussed an authentication scheme based on touchscreen. This approach has two phases, training,to model the valid subject, and an authentication phase, to make a decision if the user is legitimate or not. The authors recruited 32 subjects for collecting data. Each subject had to own his/her smartphone device, and received a \$6 gift for his/her participation. The authors gathered 50 touches in each experiment and a total of 1200 touch data sequences from each subject. The authors chose SVM \cite{SVMs} for the user authentication approach. They used Radial Basis Function (RBFN) kernel \cite{hwang1997efficient} in their approach. The authors found that their approach could achieve EER less than 10\% for all four types keystroke, slide, handwriting and pinch. However, the slide operation had achieved best accuracy, which is 0.75\%  as the number of training examples were increased from 5 to 28 .\\

\subsubsection{Multitouch Gesture-based Authentication}

Sea-Bea et al.\cite{sae2014multitouch} discussed how to enhance user authentication using multi-touch gestures. They implemented a method with several phases, which are pre-processing in which the gesture data for each touching of the screen is relabeled based on the corresponding fingertip; feature transformation to extract translation and rotation invariant features to represent the gestures; computing pairwise distance between an enrolled sample and a gesture input; and score calculation by computing dissimilarity scores from pairwise distances. The authors employed hand geometry and muscle behavior in their study.\\

The authors recruited 34 subjects to participate (24 male, 10 female) and used iPad 1 (16 GB) running iOS 3.2. to collect features. Each subject performed 22 pre-defined gestures 10 times. After every gesture, the subject filled a survey about the ease of use of the gesture before moving to the next gesture. The authors used Dynamic Time Warping (DTW) \cite{DTW} to calculate the distance between two multi-touch gestures, using three metrics for computing the cost function: Manhattan \cite{reinelt1991tsplib}, Euclidean \cite{danielsson1980euclidean}, and cosine distance \cite{smith1999integrated}. The authors allowed several single gestures such as Counter-Clock Wise rotation (CCW), pinch, drag and swipe. The verification accuracy achieved was 90\%. Average EER observed was 10.68\%. Combining CCW, Pinch, and Swipe gestures together, the approach achieved EER of 6.52\%. The system obtained the best EER of 4.03\% when using a local threshold that combined a user-defined gesture with the CCW gesture. \\

\subsubsection{Lightweight Touch Dynamics Approach}

Meng et al.\cite{meng2014design} implemented a lightweight approach, which used 8 touch gestures in order to decrease the time needed for processing data. These gestures were based on a number of touch movements, single and multiple touches and have different speeds,  durations and pressures. The authors recruited 50 subjects (26 male, 24 female) to participate in 25 sessions (1250 sessions of 120 touch events each) in 3 days. Each study used a Google/HTC Nexus One Android phone. The authors chose five classifiers, which are as Decision Tree \cite{quinlan1996bagging}, Naive Bayes \cite{rish2001empirical}, Radial Basis Function Network (RBFN) \cite{hwang1997efficient}, Back Propagation Neural Network (BPNN) \cite{paola1995detailed}, and PSO-RBFN \cite{meng2013touch}. The PSO-RBFN classifier achieved the best performance with an average EER of 2.46\% (FAR=2.55\%, FRR=2.37\%). \\

\subsubsection{Choosing Suitable Classifiers Using a Cost Metric}

Meng et al.\cite{veniamin2014effect} developed a mechanism for choosing suitable classifiers to validate users. The implemented approach extracted 6 features, which are the number of touch movements per session (NTM), the number of single-touch events per session (NST), the number of multi-touch events per session (NMT), the average duration of touch movements per session (ATTM), the average duration of single-touch events per session (ATST) and the average duration of multi-touch events per session (ATMT). The authors recruited 50 participants (27 male, 23 female) to their study for 25 sessions. Each session includes 100 touch gesture for each event. \\

The authors chose three classifiers, which were (J48) Decision Tree \cite{quinlan1996bagging}, Naive Bayes  \cite{rish2001empirical} and k-NN \cite{knn}. For five subjects, the results show that selection of the classifier may vary with user input after each experiment. For example, for user2, the J48 classifier was selected at the 1st and the 4th experiments with cost values (cost value means the relative expected cost, to evaluate different classifiers) of 1.3451 and 1.3641, respectively, while the Naive Bayes classifier was selected at the second and third experiments with cost values of 1.3221 and 1.3741, respectively. For the last experiment, k-NN outperforms the other classifiers with the cost values of 1.3321. Table \ref{table_Proc} illustrates the process of classifier selection. 

\begin{table}[!h]
	\renewcommand{\arraystretch}{1.3}
	\caption{The process of classifier selection}
	\label{table_Proc}
	\centering
	\begin {tabular}{|p{.5cm}|p{1.1cm}|p{1.1cm}|p{1.1cm}|p{1.1cm}|p{1.1cm}|}
	\hline
	\bfseries User & \bfseries 1$^{st}$ Experiment & \bfseries 2$^{nd}$ Experiment & \bfseries 3$^{rd}$ Experiment & \bfseries 4$^{th}$ Experiment & \bfseries 5$^{th}$ Experiment \\
	\hline
	1 & IBK = 1.3765 & J48 = 1.3046  & J48 = 1.2765  & J48 =1.2712  & IBK = 1.2487 \\
	
	\hline
	
	2 & J48 = 1.3451 & NBayes = 1.3221 & NBayes= 1.3741 & J48= 1.3641  & IBK = 1.3321 \\
	
	\hline
	
	3 & IBK = 1.4181 & IBK = 1.3872  & J48 = 1.3175  & NBayes = 1.3760  & J48 = 1.3029 \\
	
	\hline
	
	4 & IBK = 1.3141  & IBK = 1.3524  & J48 = 1.3121  & J48 = 1.2654  & J48 = 1.2342 \\
	
	\hline
	
	5 & J48 = 1.3742  & NBayes = 1.3891  & IBK = 1.3076  & IBK = 1.3421 & J48 = 1.2432\\
	
	\hline
\end{tabular}
\end{table}
\lipsum[0-0]

The authors calculated the accuracy for the system in terms of FAR and FRR. The proposed scheme obtained FRR between 5.87\% and 6.65\% and FAR between 6.98\% and 7.74\%. \\ 

\subsubsection{Multi-Touch Passwords (MT-Lock)}

Oakley and Bianchi \cite{oakley2012multi} investigated the effect of multiple touches when inputting passwords in an approach known as Multi-Touch Passwords for Mobile Device Access (MT-Lock). MT-Lock explores how users create and input passwords. The authors asked 10 subjects to participate in their approach (8 male, 2 female). They used an Android Nexus S smartphone (4 inch touch screen, 800x480 resolution) to gather data. Each subject typed his/her selected password six times, one time for saving the password and the rest as repetition. The authors evaluated their approach in terms of EER, and found it achieved the 8\%. \\

\subsubsection{Authentication based on Keystroke, Slide, Handwriting and Pinch}

Xu et al.\cite{xu2014towards} introduced a mechanism using touch gestures and investigate how to implement continuous validation for the legitimate user based on modeling multiple types of touch gestures. The process was based on two phases, a training phase: using labeled touch data from valid users, and an authentication phase used to compare if data come from legitimate or impostor users. The authors implemented a separation-of-concerns approach; a set of methods that 
considers separately each type of touch operations Keystroke, Slide, Handwriting and Pinch with its corresponding sequence of raw events.\\

The authors recruited 32 subjects for their experiment. They used Galaxy SII smartphone with Android OS 4.1.2 to collect features and gathered 200 touch data sequences from each subject. The authors used Support Vector Machine (SVM)\cite{SVMs} and performed a 10-fold cross validation with nine subsets for training and one for testing. This approach achieved EER below 10\% for all four types, and the best accuracy was achieved for the Slide operation that obtained EER of 0.64\%.\\

\subsubsection{Touch-based Intrusion Detection}

Damopoulos et al. \cite{damopoulos2013keyloggers} implemented an approach for iPhone devices using a touch logger that collected all touch events occurring on the screen. This approach can be used as an Intrusion Detection System (IDS) to identify possible intrusion. The authors recruited 18 subjects who owned iPhone devices. Each subject was asked to download an app that collected touch data for 24 hours. The app worked in the background and send back the log files containing touch data records. One user was referred to as the authorized user while rest of the 17 users represented attackers. The data fields in a log file were the following: Event, X, Y, Timestamps, and Attacker/Authorized. Here, Event was the type of touching of the touchscreen, X and Y were the coordinates of the touchscreen, Timestamp was the UNIX timestamp that gave the date and time a touch event occurred, and Attacker/Authorized was a boolean that indicated if the event belonged to the owner or an attacker. The authors evaluated the approach using four algorithms: Bayesian Networks, Radial Basis Function (RBF) \cite{hwang1997efficient}, K-Nearest Neighbor (kNN) \cite{knn} and Random Forest \cite{pal2005random}. The Random Forest achieved the best result in terms of accuracy, which was 99.2\%.\\

\subsubsection{Authentication by Drawing}  

Angulo and Wastlund \cite{angulo2012exploring} discussed the possibility of introducing a method for smartphone authentication based on drawing patterns on the touchscreen. This approach works on the Android platform and for testing, the authors collected patterns drawn by 32 subjects. Each subject had to draw three different patterns within a certain time and each pattern had to contain 6 dots. The implemented application collected data such as speed of finger travel from one dot to another, and the time in milliseconds from the moment the finger touched a dot to the moment the finger was pulled outside the dot area. The authors used several classifiers and metrics to compare the performance of the proposed approach: Random Forest \cite{pal2005random}, SVM \cite{SVMs}, Recursive Partitioning (RPart) classifiers \cite{zhang2011analyzing}, Manhattan distance \cite{reinelt1991tsplib}, Mahalanobis distance \cite{de2000mahalanobis} and Euclidean distance \cite{danielsson1980euclidean}. They evaluated the approach in terms of EER and the best algorithm was the Random Forest classifier with average EER of 10.3\% when an impostor already knew the user's secret pattern.\\

\subsubsection{Authenticating using Features of Host and Cloud-based IDS}

Damopoulos et al. \cite{damopoulos2014best} implemented a dynamic hybrid mobile IDS framework on iOS smartphones by introducing a method that combined features from host and cloud based anomaly detection systems. This framework used four different detection methods: SMS profiling, monitoring of applications behavior via iDMA, a touch logger profiler (iTL) and a keystroke-based authentication system. The authors collected traces from real users' profiles, and used Decision Trees \cite{quinlan1996bagging} and Random Forest \cite{pal2005random} as learning algorithms. They ran experiments to measure detection accuracy as well as CPU, memory and battery usages, and the time to detect a malicious application. The authors hosted the IDS on the iPhone 5s, and used AWS Amazon's EC2 infrastructure as the cloud service. The authors calculated the computational overhead and obtained CPU consumption between 20-45\% and memory consumption between 62-78\%. \\

Tables \ref{table_Touch}  and \ref{table_TouchB} summarize current studies that use touchscreen behavior.\\

\begin{table*}[htp]
	\renewcommand{\arraystretch}{1.3}
	\caption{Examples of Touchscreen Behavior}
	\label{table_Touch}
	\centering
	\renewcommand{\arraystretch}{1.2}
	\begin{tabular}{|p{1.7cm}|p{.9cm}|p{1.7cm}|p{2.6cm}|p{1.4cm}|p{1.4cm}|p{1cm}|p{2cm}|p{1cm}|p{.9cm}|}
		\hline
		\textbf{Study} & \textbf{Datasets} & \textbf{Classifier} & \textbf{Used Feature}& \multicolumn{6}{c|}{\textbf{RESULT}}\\
		
		\cline{5-10}
		&&&& \textbf{EER} & \textbf{FAR} & \textbf{FRR} & \textbf{Accuracy} & \textbf{Energy usage} & \textbf{Battery usage} 
		
		\\
		\hline\hline
		
		De Luca et al.\cite{de2012touch}, 2012 & 26 & DTW \cite{DTW} & X-Y coordinates, Pressure, Size, Time & 20\% & 21\% & 19\%, 77\% & - & - &- \\

		\hline
		
		Zheng et al.\cite{zheng2012you}, 2014 & 80 & 1 Class Classification \cite{1C} &Acceleration, Pressure, Size, Hold time, Inter-key & 3.58\% , 7.34\% & - & - & - & - &-  \\
		
		\hline 
		
		Zhao et al.\cite{zhao2013continuous}, 2013 & 30& $L_1$ Distance \cite{huang1997image}, $L_2$ Distance \cite{danielsson1980euclidean}, Normalized Cross Correlation \cite{lewis1995fast} &Time duration, Length of touch traces, Directions and speeds of finger movements, Tactile pressures& GTGF-A 2.62\% & - & - & - & - &-  \\
		
		\hline 
		
		Li et al. \cite{li2013unobservable}, 2013 & 75 & SVM \cite{SVMs} & Users' finger movements& - & 4\% & 4\% & - & - &-  \\
		
		\hline 
		
		Frank et al.\cite{frank2013touchalytics}, 2013& 41 & k-NN \cite{knn}, Gaussian kernel SVM \cite{rasmussen2006gaussian} & X-Y coordinates, Pressure, Speed, Length of trajectory, Inter-stroke time, Phone orientation, Finger orientation, Stroke duration & 2\%-3\% (intersession), 0\% (Intra-session) & - & - & - & - &- \\

		\hline
		
		Feng et al.\cite{feng2014tips}, 2014 & 13 & 1NN \cite{garcia2012prototype},DTW \cite{DTW} &Touch Location, Swipe/Zoom Length, Swipe/Zoom Curvature & - & - & -  & - & 88 mW & 6.2\%  \\
		
		\hline

		Shahzad et al. \cite{shahzad2013secure}, 2012 & 50 & SVDE \cite{keerthi2003asymptotic}, RBF \cite{hwang1997efficient}& Velocity magnitude, Device acceleration, Stroke time, Inter-stroke time, Stroke displacement magnitude, Stroke displacement direction, Velocity direction.  & 0.5\% &-&-&-&-&-\\

		\hline

		Feng et al.\cite{feng2012continuous}, 2012 & 40 & J48 \cite{quinlan1996bagging}, Random Forest \cite{pal2005random} ,Bayes Net \cite{friedman1997bayesian} & Gesture type, X-Y coordinates, Directions of the finger motion, Finger motion speed, Pressure, Distance between multi-touch points & 11.96\% (No gloves) 2.15\% (gloves), 4.66\% (sequence-based)  & 8.53\% (No gloves) sensors, 1.63\% (gloves), 0.13\% (sequence-based) &-&-&-&-\\
		
		\hline 
		
		Riva et al.\cite{riva2012progressive}, 2012 & 20 & SVM \cite{SVMs}, J48 \cite{quinlan1996bagging}, Linear Regression \cite{seber2012linear}& Placement of the phone, Time, Voice, Proximity of phone, Confidence & - &-&-& SVM 92.5\%, 83\%-100\% (all sensors), 76\%-99\% (accelerometers) &-&-\\
		
		\hline
		
		Saravanan et al.\cite{saravanan2014latentgesture}, 2014 & 20 & Lib-SVM \cite{chang2011libsvm}, Random Forests\cite{pal2005random} & Pressure, Timestamps, X-Y coordinates& -&- &- & 97.9\% (phone-S), 96.79\% (tablet-S), 97.78\% (phone-M) 100\% (tablet-M) &-&- \\ 
		
		\hline
			Saevanee \& Bhattarakosol\cite{saevanee2009authenticating}, 2009 & 10 & Neural Networks (PNN)\cite{PNN90}& Inter-key, Hold-time, Finger pressure & 9\% (All features), 29\% (hold-time \& inter-key) & - & - & - & - &- \\
			
			\hline 
			Burgbacher et al.\cite{burgbacher2014behavioral}, 2014  & 42 & k-NN\cite{knn} DTW \cite{DTW}& Location of the fingertip, Timestamps & 10\% -20\% & - & - & - & -&- \\

			\hline

	\end{tabular}
	
\end{table*}

\begin{table*}[htp]
	\renewcommand{\arraystretch}{1.3}
	\caption{Examples of Touchscreen Behavior}
	\label{table_TouchB}
	\centering
	\renewcommand{\arraystretch}{1.2}
	\begin{tabular}{|p{1.7cm}|p{.9cm}|p{2.5cm}|p{3cm}|p{1.7cm}|p{1.3cm}|p{1cm}|p{.5cm}|p{1cm}|p{.9cm}|}
		\hline
		\textbf{Study} & \textbf{Datasets} & \textbf{Classifier} & \textbf{Used Feature}& \multicolumn{6}{c|}{\textbf{RESULT}}\\
		
		\cline{5-10}
		&&&& \textbf{EER} & \textbf{FAR} & \textbf{FRR} & \textbf{Accuracy} & \textbf{Energy usage} & \textbf{Battery usage} 
		
		\\
		\hline\hline

			Meng et al.\cite{meng2013touch}, 2013 & 20 & J48\cite{quinlan1996bagging}, Naive Bayes \cite{rish2001empirical}, Kstar \cite{cleary1995k}, RBFN \cite{hwang1997efficient},BPNN \cite{paola1995detailed} &Gesture Type, X-Y coordinates, Time &  PSO-RBFN (RBFN + BPNN) = 2.92\% &- &-&-&-&- \\
			
			\hline

		Xu et al.\cite{xu2014towards}, 2014 & 32 & SVM \cite{SVMs},RBFN \cite{hwang1997efficient} &Time, X-Y coordinates, Pressure, Size & 10\% (All features), 0.75\% (slide)&-&-&-&-&- \\
		
		\hline 
		
		Sea-Bea et al.\cite{sae2014multitouch}, 2014 & 34 & (DTW) \cite{DTW}, Manhattan \cite{reinelt1991tsplib}, Euclidean \cite{danielsson1980euclidean}, Cosine distance \cite{smith1999integrated} & Palm movement, Fingertip movement, Dynamic fingertip & 4.03\% (all gestures) &-&-&-&-&- \\

		\hline 
		
		Meng et al.\cite{meng2014design}, 2014 & 50 & J48 \cite{quinlan1996bagging}, Naive Bayes \cite{rish2001empirical}, RBFN\cite{hwang1997efficient}, BPNN \cite{paola1995detailed},PSO-RBFN \cite{meng2013touch} & Pressure, Speed, X-Y coordinates, Gesture type& PSO-RBFN 2.46\% & 2.55\%,2.37\%&- &-&-&- \\
		
		\hline 
		
		Meng et al.\cite{veniamin2014effect}, 2014  & 50 & J48  \cite{quinlan1996bagging}, Naïve Bayes \cite{rish2001empirical},k-NN \cite{knn} & Touch movements, Single-touch events, Multi-touch events, Duration time of touch movements, Duration time of  single-touch, Duration time of multi-touch & - & 6.98\% - 7.74\% & 5.87\% -6.65\% &-&-&-\\
		
		\hline
		
		Oakley and Bianchi \cite{oakley2012multi}, 2012 & 10  & -&Raw touch events(press, motion, release) Widget and application level events (buttons press, strokes over)  &  8\% &-&-&-&-&- \\

		\hline
		Damopoulos et al. \cite{damopoulos2013keyloggers},2013& 18 & Bayesian \cite{friedman1997bayesian}, RBF \cite{hwang1997efficient}, k-NN\cite{knn} \& Random Forest\cite{pal2005random} & Gesture type, X-Y coordinates, Timestamps, Intruder/Legit& - & - & - & 99.2\% &- & - \\
		\hline 
		
		Angulo and Wastlund \cite{angulo2012exploring}, 2012 & 32 & Random Forest \cite{pal2005random}, SVM \cite{SVMs}, Recursive Partitioning (RPart) \cite{zhang2011analyzing}, the Manhattan \cite{reinelt1991tsplib}, the Mahalanobis \cite{de2000mahalanobis} \& Euclidean \cite{danielsson1980euclidean} & Finger movement, Time &10.3\% & - & - & - &-& \\
		\hline 
		
		Damopoulos et al. \cite{damopoulos2014best}, 2014 &- & J48\cite{quinlan1996bagging}\& Random Forest\cite{pal2005random} & a-SMS Profiler (Number, Timestamp, Flags, Country, Intruder/Legit)
b- iTL  (Touch Type, X-Y coordinates, Timestamps, Intruder/Legit)
c- Touchstroke (Hold-time, Inter-time, Distance, Speed,Intruder/Legit)& -&-&-& -&20-45\% & 62- 78\%\\
		\hline

	\end{tabular}
	
\end{table*}

\subsection{Gait Based Authentication}

\begin{figure}[!h]
	\centering
	\includegraphics[width=3.5in]{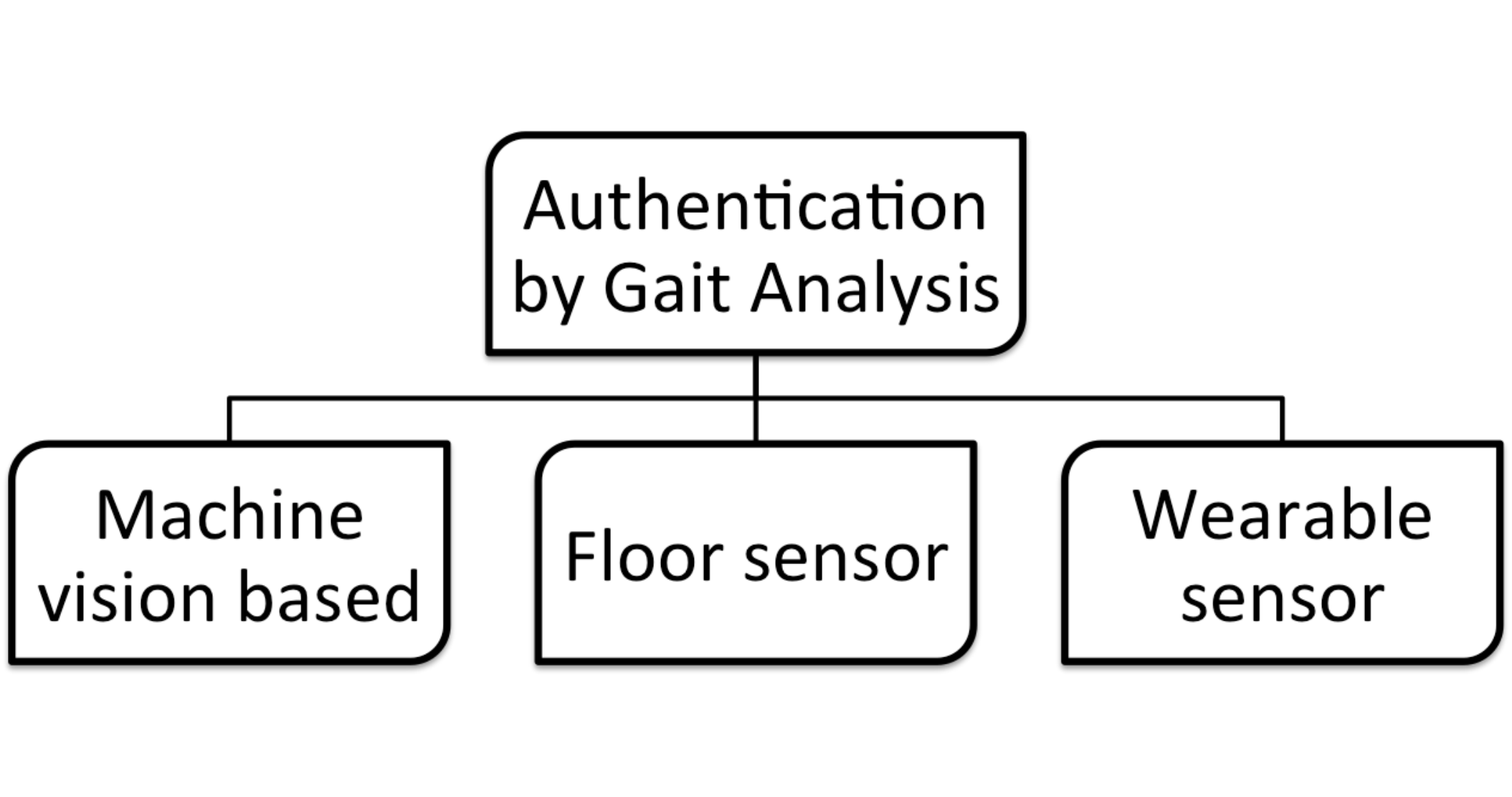}
	\caption{Gait Approaches}
	\label{fig_ga}
\end{figure}

A new approach to validating users is gait biometric. Gait biometric aims to identify and verify users' walking styles such as how a person moves at normal or fast pace \cite{gait2007survey}. Fig. \ref{fig_ga} shows main approaches to gait biometric.\\

The first two approaches that use machine vision and floor sensors are not applicable in the case of smartphones. Therefore, we concentrate on the Wearable Sensor based (WS) approach.\\

Wearable Sensors are devices worn on the bodies of subjects in order to gather information. Information can be collected using a motion recording system, which allows subjects to wear devices at any location on the human body, such as waist, belt, trouser pockets and hand. Sensors like accelerometer, speed sensors \cite{del2012speed}, gyroscope and force sensors may also be used.\\

Gait features can be extracted using cyclic and non-cyclic methods \cite{VMs}. A cyclic approach works in two steps. First, cycles in gait are identified in terms of time series. Features of these cycles are computed to extract characteristic templates for classification. This approach is easier to implement than a non-cyclic approach, which requires computing features without prior identification of cycles. A non-cyclic method chooses time intervals during walking to capture locations of sensors.\\

In this section, we discuss 13 gait-based approaches in terms of their particular methods, data acquisition, type of  classifiers used and results obtained. Table \ref{table_gait} summarizes the current studies using walking approaches.\\

\subsubsection{Gait Using Wavelet Transform and SVMs}
Hestbek et. al \cite{hgait2012} introduced a method using wearable sensors and non-cyclic feature extraction. This approach measures acceleration in 3 spatial dimensions: vertical $X$, anteroposterior $Y$ and mediolateral $Z$.  \\

The number of participants in Hestbek's study was 36, and the average length of sessions was 24 days. On each day, subjects were asked to walk up to 30 minutes in various manners of walking. One episode of walk is defined as walking the entire distance between two ends of a hallway. Each individual was asked to walk in 3 ways in sequence: 12 steps using normal gait, then 16 steps at a quick pace and finally 12 steps at normal walking speed. The authors used Wavelet Transforms (WT) \cite{WT} to process signal information. The approach divided a signal into low-pass and high-pass information to represent coarse and clear variations based on time domain. The authors used DWT \cite{DTW} to represent and transform the spatial data into Approximation signals and Detail coefficients.\\

The authors used SVMs\cite{SVMs} for classification of the data into two groups, authorized and unauthorized subjects. The authors experimented with 9 different feature sets, which combine Bark-Frequency Cepstral Coefficients (BFCC) and Standard Deviation (SD), and approximation and detail coefficients of gait templates. Feature sets 1 and 2 were previously used by \cite{VMs} while the rest of the sets were extension by Hestbek et al. with Discrete Wavelet Transform (DWT) \cite{hgait2012}. Table \ref{table_F1} shows all 9 feature sets.\\ 
\begin{table}[h]
	\renewcommand{\arraystretch}{1.3}
	\caption{Feature Sets}
	\label{table_F1}
	\centering
	\begin{tabular}{|c|c|c|c|}
		\hline
		\bfseries Set & \bfseries Feature & \bfseries Acceleration & \bfseries Signal Coefficients \\
		\hline\hline
		1 & BFCC & a,v  & [$P^1$]\\
		\hline 
		2 & BFCC & a,v  & [$P^6$], [$T^4--T^6$]\\
		\hline 
		3 & BFCC,SD & [a,v],[v] & [$P^1$], [$P^6, T^4--T^6$] \\
		\hline 
		4 & BFCC,SD & [a,v],[v]  & [$P^1$], [$P^6$, $T^4,T^5$] \\
		\hline 
		5 & BFCC,SD & [a,v],[v]  & [$P^1,P^2$], [$P^6, T^4, T^5$] \\
		\hline 
		6 & BFCC,SD & [a],[v]  & [$P^1$], [$P^6, T^4--T^6$ ] \\
		\hline 
		7 & BFCC,SD & [b],[v] & [$P^1$], [$P^6, T^4--T^6$] \\
		\hline 
		8 & BFCC,SD & [c],[v]  & [$P^1$], [$P^6, T^4--T^6$] \\
		\hline 
		9 & BFCC,SD & [v],[v]  & [$P^1$], [$P^6, T^4--T^6$] \\
		\hline
	\end{tabular}
\end{table}
\lipsum[0-0]

Hestbek et al. divided their experiments into training and testing phases. The authors used three evaluation metrics, which are False Match Rate (FMR), False Non-Match Rate (FNMR), and Half Total Error Rate (HTER). After experiments with various interpolation rates, segmentation lengths and different wavelets, the authors found that the Haar Transfer procedure \cite{stankovic2003haar} had the best accuracy. As a result, the authors used Haar Transform in further experiments. They also found the best interpolation rate is 400Hz and the best segment length was 10 seconds. The best feature set obtained FMR = 3.48\%, FNMR = 34.36\% and HTER = 18.92. Values of FNMR were improved using a voting method \cite{VMs}, which reduces the number of legitimate persons who were incorrectly rejected. The values obtained using feature set 6 were FMR = 9.82\%, FNMR = 10.45\% and HTER = 10.14. Using the voting method, the approach achieved FMR = 9.82\%, FNMR = 10.45\% and HTER = 10.14. Table \ref{table_G2} provides a comparison of all 9 feature sets in terms of FMR, FNMR and HTER. 

\begin{table}[h]
	\renewcommand{\arraystretch}{1.3}
	\caption{Evaluation Results of Recognition using at 400 Hz and 10 seconds}
	\label{table_G2}
	\centering
	\begin{tabular}{|c|c|c|c|}
		\hline
		\bfseries Set & \bfseries FMR & \bfseries FNMR & \bfseries HTER \\
		\hline\hline
		1 & 1.58\% & 40.56\% & 21.07\\
		\hline 
		2 & 1.42\% & 44.19\% & 22.81\\
		\hline 
		3 & 2.29\% & 39.16\% & 20.73\\
		\hline 
		4 & 2.19\% & 39.27\% & 20.74\\
		\hline 
		5 & 1.92\% & 40.06\% & 21.00\\
		\hline 
		6 & 3.48\% & 34.36\% & 18.92\\
		\hline 
		7 & 3.38\% & 63.80\% & 33.90\\
		\hline 
		8 & 2.94\% & 62.85\% & 32.90\\
		\hline 
		9 & 3.99\% & 48.44\% & 26.21 \\
		
		\hline
	\end{tabular}
\end{table}
\lipsum[0-0]

\subsubsection{Gait Authentication and Identification using Wearable Accelerometer Sensor}

Gafurov et al.\cite{gait2007} used an accelerometer sensor attached to the trouser pocket of the subject to gather gait features. The authors used cyclic features. The goal of this study is to measure performance in identification and authentication when a user walks while he/she may be carrying a backpack by calculating an appropriate metric. In order to evaluate the performance of the proposed approach from two perspectives, authentication and identification, the authors used four methods, which are absolute distance, correlation, histogram and higher order moments. When calculating absolute distance and correlation, they computed an average cycle for each subject as a feature vector. The histogram feature uses n-bin (e.g., n=100) and uses it as a feature vector. Computing skewness and kurtosis was performed using higher order moments.\\

Gafurov et al. had 50 participants, and each participant had to walk 4 normal walking episodes without a backpack and then 2 walking episodes carrying a backpack. They collected 300 gait samples. The authors used Manhattan distance \cite{reinelt1991tsplib} and Euclidean distance \cite{danielsson1980euclidean} to compute the dissimilarity score between feature vectors. For authentication, the use of absolute distance achieved the best performance  with values in the histogram and higher order moments used in identification become very low. The performance deteriorated from about 7.3\% to about 9.3\% in samples without backpack to with backpack. For identification, recognition rate dropped slightly from 86.3\% to 86.2\%.\\

\subsubsection{Extracting Gait Cycle using Piecewise Linear Approximation}

Muaaz and Mayrhofer \cite{muaaz2013analysis} implemented a technique for extracting gait cycle using Piecewise Linear Approximation (PLA). The authors introduced two approaches to classify gait features: using a pre-computed data matrix and using a pre-computed kernel matrix to help build an elastic similarity measure. This elastic measure uses a function called Gaussian Dynamic Time Warp (GDTW) kernel \cite{keogh2005exact}. \\

The authors asked 51 subjects to participate in this approach. To collect data, the subjects attached Personal Mobile Devices (PMDs) on the right hand of the hip. The authors used two classifiers in two stages, which are Dynamic Time Warping (DTW)\cite{DTW} and Gaussian Dynamic Time Warp (GDTW) kernel \cite{keogh2005exact}. 
The authors evaluated their approach with a method presented in \cite{muaaz2012influence} in terms of EER in two ways: using PLA and not using PLA. Results obtained without PLA are slightly higher than the ones achieved by introducing PLA. For same day experiments, EER value with PLA is 22.49\% and without PLA is 16.26\%. For different day experiments, with PLA the EER is 33.3\% while without PLA it is 28.21\%). \\

\subsubsection{Using SVMs and HMMs for accelerometer-based biometric gait recognition}

Nickel et al.\cite{N11B} used fixed-length  time  segments for extracting features. The authors used SVMs\cite{SVMs} and HMMs\cite{rabiner1989tutorial} for classifying to distinguish users. 

Nickel et al.\cite{N11B} asked 36 subjects to participate in their study in two sessions. Each subject attached the phone to his/her right hip and took 12 normal walks, 16 fast walks and finally 12 normal walks. Subjects had to walk on flat carpet. There were three preprocessing steps. The first step, the mean sampling rate of the database was 127.46 data values per second (min= 109.19, max= 128.88). Then, the sampled data was interpolated at 50, 100 and 200 data values per second. Finally, the data was segmented into 3 seconds, 5 seconds and 7.5 seconds with  an overlap of 50\%.\\

The authors used two machine learning algorithms, which are the SVM implementation LIB-SVM \cite{chang2011libsvm} and HMM implementation in the Hidden Markov Model Toolkit (HTK)\cite{young2006htk}. Normal walking obtained TER of 29.20\% while for fast walking, it was 30.87\%. A voting method \cite{VMs} improved the results for HMM EER to 15.77\% for normal walk and to 14.39\% for fast walk. For SVM, the TER is  20.01\% for normal walk, while the EER of HMMs is 12.63\%. \\

\subsubsection{Walk Authentication using k-NN algorithm}

Nickel et al.\cite{nickel2012authentication} implemented an approach for classification of gaits using the k-NN algorithm \cite{knn} and compared this approach with theirs previous approach \cite{N11B}. In this new approach, Nickel et al. used the same dataset that was used in \cite{N11B} with 36 subjects. The mean sampling rate is 127 data values per second. The data was divided into segments of fixed time length, which was between 3 and 7.5s with an overlap of 50\%. Euclidean distance \cite{danielsson1980euclidean} was used to compute distance between the probe vector and stored instances. The k-NN implementation of the WEKA \cite{WEKA} library was used. 
The authors compared the results for this approach with their approach in \cite{N11B}. The lowest EER/HTER was between 8.24 and 8.85. A walking duration of five minutes for the training phase was identified to be best for all algorithms. Based on the error rates, no clearly best stochastic classifier could be identified. A disadvantage of SVMs was that outliers occurred in training, resulting in very high FNMRs. If this happens during an experiment, the used classifier was unsuitable and the experiment has to be repeated, resulting in lower usability. k-NN gives the lowest HTER before and after voting \cite{VMs}. \\

\subsubsection{Using Majority Voting Module and Cyclic Rotation Metric Module for Gait Recognition}

Nickel et al.\cite{neer21} implemented a method based on salience vectors, which had a minimum height and distance of half the estimated cycle length. The data obtained from 48 subjects, where each participant had to walk on several different days. The phone was inside a pouch which was attached to the right hip of the subject. There were two sessions. In the enrollment session, each participant had to walk straight for 10 seconds on a flat floor. In the second session, the participant had to walk using a predefined route three times. For each of the 48 subjects, they obtained 28 data sets in each session, 2,688 in total. The data was linearly interpolated to a fixed rate of 25, 50, and 100 samples per second, and the interpolated data were divided into segments of 2, 3, and 4 s, with an overlap of 50\%.\\

The authors introduced two methods Majority Voting Module; to calculate the reference cycles among enrollment phase using DTW \cite{DTW} and Cyclic Rotation Metric Module to compare two sets using Manhattan distance \cite{reinelt1991tsplib} and DTW. For a module using Majority Voting, the approach obtained EER of 28\% while it obtained 21.7\% using cyclic rotation metric.\\  

\subsubsection{Classifying accelerometer data using Hidden Markov Models}

Nickel et al.\cite{nickel2013classifying} introduced another classifier to compare with previous works. The author used the data collection that was used in \cite{neer21}.\\

Hidden Markov Models (HMMs) \cite{rabiner1989tutorial} were used in this approach. HMMs were trained using data from the subjects. Given a probe feature vector, they obtained for each model the probability that this model represented the probe data. Classification was based on the difference between these probabilities and a decision threshold. Calculated without voting  and using segments size of 4000, the EER was between 15.77\% and 7.45\%. When a segment size of 2s was used, the EER was 7.33\%. With same day evaluation, the EER reached 0.71\%. On the other hand, once the authors used voting method \cite{VMs}, the approach obtained EER of 5.81\%. \\

\subsubsection{Cyclic Recognition using Voting}

Derawi and Bours \cite{derawi2013gait} used Samsung Nexus S (Gingerbread operating system) and implemented an application for collecting data. The authors used 25 subjects for their experiments. All targets had to attach the phone to pockets of their trousers. Derawi and Bours divided participants into two groups. The first group consisted of 5 subjects, who had to walk with three different speeds and represented authorized people. The second group contained 20 subjects for testing purposes, representing impostors. All participants had to walk 30 meters. Each subject walked 5 times at normal speed, 5 times at slow speed and 5 times at fast speed. A total of 375 data samples (300 data samples are impostors and 75 are genuine) were collected.\\ 

The authors used Euclidean distance \cite{danielsson1980euclidean}, Manhattan distance \cite{reinelt1991tsplib}, Dynamic Time Warping (DTW) distance \cite{DTW}, and the Cross DTW (CDM) as distance metrics. The authors also used several classifiers, which are 
Bayes Net \cite{friedman1997bayesian}, Lib-SVM\cite{chang2011libsvm}, Logistic Model Trees (LMT) \cite{landwehr2005logistic}, Multilayer Percepton (MLP) \cite{pal1992multilayer}, Naive Bayes \cite{rish2001empirical}, (RBFN) \cite{hwang1997efficient} and Random Tree \cite{pal2005random}. The best result was obtained with LibSVM with an accuracy of 99.6\%. Next, the RBFN achieved 98.9\%.  \\

Global cross-validation was performed by merging the data from all 25 subjects. The results indicate how different normal, fast and slow walks are within a group of users. The LibSVM and LMT \cite{landwehr2005logistic} slightly outperformed the other methods with recognition rates of 87.6\% and 86.7\%, respectively. The correct classification probability of a walk can be estimated from the correct classification probability of the cycles, by simply using majority voting over the cycles in a walk. Assume that at least 9 cycles are detected in a walk, and assume that the correct classification probability of a single cycle was as low as 80\%. Even under these assumptions, a walk was classified correctly in over 98\% of the cases. The data samples in the experiment represented a walk of approximately 30 meters. For all these data samples, the assumption was that at least 9 cycles detected were always correct.\\ 

\subsubsection{Recognition Gait Patterns with Accelerometers}

Mantyjarvi et al.\cite{mantyjarvi2005identifying} introduced a scheme to distinguish and identify users based on walking patterns using acceleration signals. They recruited 36 subjects, 19 males and 17 females, to walk at three different speeds, fast, normal and slow. They had to walk approximately 20 meters at the three speeds. The data was collected over a period of 5 days. Each subject had to wear the accelerometer on his/her belt in the back. The accelerometer signals were recorded at 256 Hz sampling frequency using a laptop computer equipped with National Instruments DAQ 1200 card.\\

The authors compared four elements computed from the signals, which are signal correlation, FFT coefficients (signal values in the frequency domain), histogram and higher order moments. The best value of EER achieved by signal correlation method was 7\%, with FFT coefficients 10\%, with higher order moments 18\% and finally with histogram 19\%. \\

\subsubsection{User Authentication at Low-grade Acceleration} 

Derawi et al.\cite{derawi2010unobtrusive} presented a method that differs from other approaches in terms of the level of gathered data. The method focused on collecting data at low-grade acceleration, while other studies concentrated in high-grade acceleration. \\

Data was collected using Google G1 mobile phones. The authors asked each of 52 participants to attach the device on the right side of the hip. Each participant had to walk 37 meters on flat carpet. A participant waited for 2 seconds at the end, and then turned around, waited 2 more seconds, then walked back the same distance. \\

The cycle length was extracted from the data and the range was between 40-60 samples. To calculate the average cycle, the authors eliminated all irregular cycles using Dynamic Time Warping (DTW)\cite{DTW}.  \\

The authors compared their approach with Holien's approach \cite{fuxreiter2010modular} in terms of EER. Derawi's approach obtained EER of 20.1\%  while Holien's approach obtained 12.9\%. In Derawi's approach, they used 40-50 samples per second whereas Holien's approach used 100 samples per second.\\

\subsubsection{Recognition with Time-Delay Embeddings}

Frank et al.\cite{frank2010activity} introduced an approach for gait recognition incorporating feature extraction using time-delay embedding and supervised learning. Time- delay embedding is a technique from analysis of nonlinear time series data, that aims to reconstruct the state of an unknown dynamical system from observations taken over time \cite{kantz2004nonlinear}. The authors also used an algorithm to perform classification of activities in real-time. This approach aims to decrease the computational complexity and memory usage in conducting activity recognition. The main idea is to extract features by employing time-delay embeddings and a noise reduction phase. The resulting features from these processes are used as input to the classifier. \\

The authors used HTC G1 phone to collect data from 25 participants. Each participant had to perform activities such as walking, lingering, running, up-stairs and downstairs. The accelerometer data was sampled at 512Hz, which decimated to 32Hz. The authors used an SVM classifier \cite{SVMs} to incorporate the features along with barometric pressure. The authors calculated the accuracy with and without using time-embeddings in the dataset that they collected. The approach achieved accuracy of 100\% on a noisy dataset, with average accuracy of approximately 85.48\%. \\

\subsubsection{Using Time and Frequency Domains}

Thang et al.\cite{thang2012gait} investigated how to identify the legitimate user by using data from the accelerometer sensor. The first approach collected data in the time domain to evaluate the similarity score whereas the second method used the frequency domain. Thang et al. used one device; the Google Android HTC Nexus. The participants had to attach the device to their trouser pockets. The sample rate was 27 Hz, linearly interpolated to 32 Hz. They divided data into windows of 8 consecutive gait cycles. The size of window frames was invariant with 50\% overlap. Each volunteer was asked to walk as naturally as possible for 12 walks with 36 seconds for each lap. \\

The authors used DTW \cite{DTW} for evaluating similarity in the time domain and used SVM \cite{SVMs} for classification in the frequency domain. The accuracy achieved in the time domain was 79.1\% while the accuracy for frequency domain was 92.7\%. \\

\subsubsection{Using Wireless Acceleration Sensors}

Choi et al.\cite{choi2014biometric} introduced 6 gait metrics to observe and capture patterns of walking. The authors used sensors that measure the variance in walking in different directions. These sensors are lightweight wireless accelerometer sensors. The metrics are derived from the rates of change in acceleration data. In particular, the authors chose a sensor called Shimmer \footnote{http://www.shimmer-research.com}, which is a wearable device with a tri-axial MMA 7361 accelerometer. The authors asked 10 participants, 5 male and 5 female, to wear this sensor in their back, wear tennis shoes and walk at normal speed on a hard floor for approximately for 30 seconds. The system applied the k-NN algorithm \cite{knn} using all six new metrics, with normalized Euclidean distance, to find the closest match between a test example and training examples. The approach identified the specific user among a list of 10 test subjects. Thus, the scheme is quite effective in extracting distinctive gait characteristics.

Table \ref{table_gait} summarizes current studies that use gait behavior.\\

\begin{table*}[htp]
	\renewcommand{\arraystretch}{1.3}
	\caption{Examples of Gait Biometrics}
	\label{table_gait}
	\centering
	\renewcommand{\arraystretch}{1.2}
	\begin{tabular}{|p{2cm}|p{.9cm}|p{2.3cm}|p{4cm}|p{2.3cm}|p{1cm}|p{0.8cm}|p{0.8cm}|p{1.5cm}|}
		\hline
		\textbf{Study} & \textbf{Dataset} & \textbf{Classifier/ Distance Metrics} & \textbf{Feature Set}& \multicolumn{5}{c|}{\textbf{RESULT}}\\
		
		\cline{5-9}
		&&&& \textbf{EER} & \textbf{FMR} & \textbf{FNMR} & \textbf{HTER} & \textbf{ACCUARCY}  
		
		\\
		\hline\hline

		Hestbek et. al \cite{hgait2012}, 2012 & 36  & DWT \cite{DTW} and SVMs\cite{SVMs} & Raw acceleration data, Variability of acceleration in lateral, Vertical and anterior/posterior directions, Threshold value of acceleration variability, Static and dynamic portions of gait cycles &- &  9.82\%, & 10.45\% & 10.14 &- \\

		\hline 
		
		Gafurov et al.\cite{gait2007}, 2007 & 50 & Manhattan  \cite{reinelt1991tsplib} and Euclidean distances \cite{danielsson1980euclidean} &Time domain features (value between -1 and +1), Interpolation rate 32Hz &7.3\% (absolute distance) \& 9.3\% & - & - & - & 86.3\%  \\

		\hline 
		
		Muaaz and Mayrhofer \cite{muaaz2013analysis}, 2013 &  51  & DTW\cite{DTW} and GDTW \cite{keogh2005exact} & Three dimensional (X, Y, Z), Sampling frequency 40-50 Hz &  22.49\% (PLA), 16.26\%  (NO PLA), 33.3\%  (PLA \ different day)  28.21\%(PLA\ different day) & - & - & - & - \\

		\hline
		
		Nickel et al.\cite{N11B}, 2011 & 36 & LIB-SVM \cite{chang2011libsvm} and HMMs\cite{rabiner1989tutorial} & MFCC Mel-Frequency Cepstral Coefficients [Max frequency= 10Hz], Bark-frequency Cepstral coefficients [Max frequency= 7.5Hz,8.75Hz]& 15.77\% (normal), 14.39\% (fast) &- & - &- &-   \\  
		\hline
		
		Nickel et al.\cite{nickel2012authentication}, 2012
		&  36 from \cite{N11B} & Euclidean distance \cite{danielsson1980euclidean} & X-Y-Z directions, Bin Relative histogram distribution in linear spaced bins between the minimum and the maximum acceleration in the segment&  8.24 & - & - & 8.85\% & -  \\
		
		\hline
		
		Nickel et al.\cite{nickel2013classifying}, 2013
		& 48 from \cite{neer21}  & HMMs\cite{rabiner1989tutorial} & Interpolation rate 25, 50, and 100 samples per second, Segment of interpolated data: 2000, 3000, and 4000 ms, Mel-frequency (MFCCs) and Bark-Frequency Cepstral Coefficients (BFCCs)& 7.33\% (same day, segment size 2s), 5.81\% (voting method) & - & - & - & - \\

		\hline
		
		Derawi and Bours \cite{derawi2013gait}, 2013 & 25  & Bayes Net \cite{friedman1997bayesian}, Lib-SVM\cite{chang2011libsvm}, Logistic Model Trees (LMT) \cite{landwehr2005logistic}, MLP \cite{pal1992multilayer}, Naive Bayes \cite{rish2001empirical}, RBFN \cite{hwang1997efficient} \& Random Tree \cite{pal2005random}) & X-Y-Z directions, Linear time interpolation, the average cycle length& - & - & - & - & 99.6\%(Lib-SVM), 98.9\% (RBFN)\\
		
		\hline

		Mantyjarvi et al.\cite{mantyjarvi2005identifying}, 2005 & 36 & - & X-Y-Z axis accelerometer, Frequency domain using 256 sample frame with 128 samples, 10-bin histograms normalized by the length of the data are composed for x and z acceleration signals
 & (signal correlation) 7\%, (FFT coefficients) 10\% & - &-&-& -  \\

		\hline
		
		Derawi et al.\cite{derawi2010unobtrusive}, 2010
		& 52  & DTW\cite{DTW} & Time Interpolation, Average cycle length & 20.1\% & - &- &- &- \\

		\hline 
		
		Frank et al.\cite{frank2010activity}, 2010 & 25 & SVM  \cite{SVMs} & X-Y-Z axis accelerometer, Pressure sensor, Frequency domain features (32Hz) &- &- &- & - & 85.48\% \\

		\hline 
		
		Thang et al.\cite{thang2012gait},2012 & 11 & DTW \cite{DTW}, SVM \cite{SVMs} & Multi-level wavelet decompositions, Time domain features, Frequency domain features (32Hz) & - &  - & - &- & 79.1\% (time domain), 92.7\% (frequency domain) \\

		\hline 
		
		Choi et al.\cite{choi2014biometric}, 2014 & 10  & k-NN \cite{knn}, Euclidean distance \cite{danielsson1980euclidean} & A) Dynamic portion of the acceleration data: Vertical acceleration metric, Lateral acceleration metric, Anterior and posterior acceleration metric. B) Standard deviations of the three dimensions: Vertical acceleration metric, Lateral acceleration metric, Anterior and posterior acceleration metric. & - &-&-&- &- \\
		
		\hline

	\end{tabular}
	
\end{table*}

\subsection{Other Behavioral biometrics} 

There are several other forms of behavioral biometrics such as voice, signature and Profiling behaviors that we have not discussed so far. In this section, we introduce these patterns and briefly outline the state-of-the-art for each.\\

\subsubsection*{Signature Behavior}
Signature recognition can be thought of as a behavioral biometric although the measurement and analysis of signatures and the activities that generate signatures are physical \cite{faundez2005signature}. Efforts at signature recognition can be categorized into two types, which are static (writing a signature on regular paper) and dynamic (writing a signature on a device such as tablets and smartphone). To recognize the user, there are many features can be extracted such as writing pressure, pen up and down directions, azimuth, inclination and spatial X and Y coordinates at time t. Table \ref{table_Signature} summarizes recent published studies in this field with respect to methodology, data collection, classifiers used and results.  \\

\begin{table*}[!h]
	\renewcommand{\arraystretch}{1.3}
	\caption{Example of Signature Behaviors on Mobile Devices}
	\label{table_Signature}
	\centering
	\renewcommand{\arraystretch}{1.2}
	\begin{tabular}{|p{2cm}|p{4cm}|p{1.5cm}|p{2.5cm}|p{1.8cm}|p{.5cm}|p{.5cm}|p{2cm}|}
		\hline
		\textbf{Study} & {\textbf{Idea}} & \textbf{Dataset} & \textbf{Classifier/ Distance Metrics} & \multicolumn{4}{c|}{\textbf{RESULT}}\\
		
		\cline{5-8}
		&&&& \textbf{EER} & \textbf{FAR} & \textbf{FRR} & \textbf{Accuracy} \\
		\hline\hline
		
		Narayanaswamy\cite{narayanaswamy1999user},1999 & Signature verification based on a specifically designed user interface in both hardware and software. &	542 authorized and 325 impostor	& HMMs \cite{rabiner1989tutorial} & 3\% & - & - & -  \\
		\hline
		
		Clarke and Mekala\cite{clarke2007application}, 2007 & Tested dynamic signatures by typing common words & 20 &- & - & 0\% & 1.2\% & - \\
		\hline

		Martinez-Diaz \cite{martinez2008towards}, 2008 & Used dynamic features (time, speed and acceleration, direction and geometry) with PDA and pen tablet devices to produce a genuine signature & 120 & Combining PDA features with HMM \cite{rabiner1989tutorial}  & 4\% & - & - & - \\  
		\hline

		Houmani et al.\cite{houmani2012biosecure}, 2012 & Evaluated several online signature approaches in two ways: Studying the effect of digital devices and measuring the influence  of information content in signatures. & 382 & DTW \cite{DTW}, Mahalanobis \cite{de2000mahalanobis} and Euclidean distances \cite{danielsson1980euclidean} and HMMs \cite{rabiner1989tutorial} & T1: 2.2\% (forgeries), 0.51\% (random forgeries). T2: 4.97\% (skilled forgeries),0.55\% (random forgeries) & - & - & -  \\
		\hline
		
		Saevanee \cite{saevanee2011sms}, 2011 & A feasibility study based on writing vocabulary words and style of SMS messages & 30 &
		RBFN \cite{hwang1997efficient} & 24\% & - & - &-  \\ 
		\hline

		Blanco et al.\cite{blanco2012handwritten}, 2012 & Studied the effect of several parameters like screen size, operative system and interoperability between the devices & 11 subjects and 8 mobile devices &  DTW-based signature recognition algorithm \cite{pascual2009practical} & Best 0.17\%  (stylus), 0.29\% (fingers), Worst 3.48\% (Asus ) & - & - & -  \\
		\hline
		
		Blanco-Gonzalo \cite{blanco2013performance}, 2013 & Matched with handwritten signature to recognize users on different devices & 43  & DTW  \cite{pascual2009practical} & Best 0.19\% (iPad), Worst 1.45\% Intuos) &-&-&- \\
		\hline

		Sae-Bae\cite{sae2014online}, 2014 & Developed a method based on online signature by drawn with a fingertip & 100 from \cite{ortega2003mcyt}, and 94 from \cite{kholmatov2009susig} &  DTW\cite{DTW}, HMMs \cite{rabiner1989tutorial} & 0.35\% (training set was 20 times) &-&-&-\\
		\hline
		
		Nguyen et al. \cite{van2014finger}, 2014 & Investigated the use of online finger-drawn PIN on a touch interface & 40 subjects and 2400 imitating samples from two attack (PIN attack, Imitating Attack) & DTW \cite{DTW} algorithm & PIN: 6.7\%, Imitation: 9.9\% &-&-&- \\
		\hline
		
		Ketabdar \cite{ketabdar2012magnetic}, 2012 & Implemented MagiSign 3D signature detector &- &  DTW \cite{DTW} & -&-&-&- \\
		
		\hline
		
		Roshandel et al.\cite{roshandel2014multi}, 2012 & Introduced Pingu, based on a 3D signature using several sensors &  24  & J48 \cite{quinlan1996bagging}, MLP \cite{pal1992multilayer}, Naïve Bayes \cite{rish2001empirical} and SVM \cite{SVMs} &-&-&-& Sig. in the air: SVM 99.1667\%, Sig. with table SVM 99.4366\% \\
		
		\hline
	\end{tabular}
	
\end{table*} 

\subsubsection*{Voice Behavior}

Another example of behavioral biometric is voice that can be used to identify a user based on his/her manner and pattern of  speaking. The speaking pattern for each person is likely to be different, based on accent, inflectional patterns, cultural background, etc. There are two types of features that can be used in voice biometric, which are text-dependent (the text to be spoken for both enrollment and authentication phases must be identical) and text-independent (no restriction on what is spoken in enrollment and authentication phases). Table \ref{table_Voice} summarizes recent published studies in this field with respect to methodology, data collection, classifiers used and results. \\

\begin{table*}[!h]
	\renewcommand{\arraystretch}{1.3}
	\caption{Examples of Applications of Voice Biometric}
	\label{table_Voice}
	\centering
	\renewcommand{\arraystretch}{1.2}
	\begin{tabular}{|c|p{4cm}|p{2cm}|p{2.5cm}|p{2cm}|p{1cm}|p{1cm}|p{2cm}|}
		\hline
		\textbf{Study} & {\textbf{Idea}} & \textbf{Dataset} & \textbf{Classifier} & \multicolumn{3}{c|}{\textbf{RESULT}}\\
		
		\cline{5-7}
		&&&& \textbf{EER} & \textbf{Precision} & \textbf{Recall}\\
		\hline\hline

		Das et al.\cite{das2008multilingual}, 2008 & Implemented Compressed Feature Dynamics (CFD) to capture speaker's identity based on speech dynamics in spoken passwords. & 79  & DTW\cite{DTW} & 0.47\% &- &- \\
		\hline
		
		Miluzzo et al.\cite{miluzzo2010darwin}, 2010 & Implemented Darwin which is applicable to several sensing apps like a speaker recognition that offers security for the content of conversations and raw data between phones. & 8  & GMM algorithm \cite{reynolds2000speaker} &-&-&- \\
		\hline
		
		Lu et al.\cite{lu2011speakersense}, 2011 & Proposed SpeakerSense that uses a heterogeneous multi-processor (HMP) to collect data via phone calls &  17 & GMM\cite{reynolds2000speaker} &- & 93\% & 85\%  \\
		\hline

		Kunz et al.\cite{kunz2011continuous}, 2011 & Introduced a method for continuous speaker verification during an ongoing phone call. & 14  & HMMs \cite{rabiner1989tutorial} & 15\% (segment length 2seconds) & -&- \\
		\hline

		Baloul et al.\cite{baloul2012challenge}, 2012 & Implemented a scheme  to defend against replay attacks & 16  from \cite{CMUDB} & VQ (vector quantization) method \cite{soong1987report} based on LBG algorithm \cite{linde1980algorithm} & 0.83\% & - &- \\

		\hline
	\end{tabular}
	
\end{table*}

\subsubsection*{Behavioral Profiling}

Behavioral profiling is another example of a behavioral biometric. It can be used to identify an individual based on interaction with various digital services and applications. Behavioral profiling can be divided into two categories, network-based (monitoring users' behavior in regards to service providers, Wi-Fi networks, Bluetooth patterns, etc.), and host-based approach (how the user uses different applications at different times and different locations). There are two levels of application profiling, which are: collecting generic application-level information e.g., application name, date and time of usage) and application-specific information like specific of voice calls and text messages including phone numbers and actual contents. Table \ref{table_BB} summarizes recently published studies in this field with respect to methodology, data collection, type of classifier and results. \\

\begin{table*}[!h]
	\renewcommand{\arraystretch}{1.3}
	\caption{Example of Behavioral Profiling}
	\label{table_BB}
	\centering
	\renewcommand{\arraystretch}{1.2}
	\begin{tabular}{|p{2.5cm}|p{4.5cm}|p{1.5cm}|p{2.5cm}|p{2cm}|p{1.5cm}|p{1.5cm}|p{1cm}|}
		\hline
		\textbf{Study} & {\textbf{Idea}} & \textbf{Dataset} & \textbf{Classifier / Static metrics} & \multicolumn{3}{c|}{\textbf{RESULT}}\\
		
		\cline{5-7}
		&&&& \textbf{EER} & \textbf{FAR} & \textbf{FPR}   \\
		\hline\hline
		
		Hall et al.\cite{hall2005anomaly}, 2005 & Investigated the feasibility of using profiles at application level, which are based on the mobility patterns using Anomaly based Intrusion Detection &  50 & Instance based learning (IBL) \cite{aha1991instance} & - & - & 100\% \\

		\hline 
		
		Li et al. \cite{li2010behaviour}, 2010 & Discussed the feasibility identifying the users based on application and network level behaviors (applications, Bluetooth scanning, Charging device, On/Off, SMS and voice) & 94 from \cite{eagle2009inferring} & Neural networks \cite{bishop1995neural}& Telephony 13.5\%, Device Usage:35.1\% and Bluetooth Scan 35.7\% &-&- \\
		\hline

		Li et al.\cite{li2014active}, 2014 & Implemented Transparent Authentication System (TAS) based on historical application usage (location and calling number) to verify mobile users. & 76 from \cite{eagle2009inferring} & RBFN \cite{hwang1997efficient}, FF-MLP\cite{bishop1995neural} and a rule-based approach\cite{brill1992simple} & 9.8\% & - & -\\ 
		\hline
		
		Bassu et al.\cite{bassu2013new}, 2013 & Designed and Implemented an active authentication approach using mobile usage context (app usage, location, time, bandwidth usage and human device interaction)  & - &  Naive Bayes model\cite{rish2001empirical} & - & - & - \\
		\hline
		
		Khan and Hengartner \cite{khan2014towards}, 2014 & Introduced application-centric approach (implement four apps: browser, launcher, maps and comics) to collect data and make authentication decision instead of Device-centric approach & 61 & SVM\cite{SVMs} & - & 5\%, 6\%, 3\% \& 9\% for launcher, browser, maps, and comics applications, respectively & -\\
		\hline

		Hayashi et al.\cite{hayashi2012goldilocks}, 2012 & Introduced an all-or-nothing access model that uses authentication when a sensitive application is launched by a user. The authors investigated how the participants can share their phones with other in several scenarios (always available, split and after unlock for both tablets and phones) & 14  & Calculate the median for all activities & - & - & - \\
		\hline
		
		Seifert et al. \cite{seifert2010treasurephone}, 2010 & Implemented TreasurePhone to using basic functions of  standard mobile phones, such as calling, SMS, address books, cameras usage and photo viewing & 20  & Calculated median, Standard deviation and the average & - & - & - \\
		\hline 
		
		Shi et al.\cite{shi2011implicit}, 2011 & Introduced a method based on user habits to compute scores, positive or negative. If the score is close to the lower limit, the user is authenticated & 50  & Calculated probability using Expectation Maximization Algorithm \cite{EM} &- & - & - \\ 
		\hline 
		Papamartzivanos et al.\cite{papamartzivanos2014cloud}, 2014 & Implemented a host and cloud approach to notify the user about misbehaving apps & - & - &- & - & - \\

		\hline
	\end{tabular}
	
\end{table*}

\subsection{ How Users View Authentication} 

To build authentication techniques that are going to be actually used, it is necessary to understand users' perspectives on the nature and use of a proposed authentication system. This section focuses on polls and surveys, which have been conducted to elicit smartphone owners' viewpoints.\\

\subsubsection{The risk of insider attack}
Muslukhov et al.\cite{muslukhov2013know} found that many users  worry about someone else accessing their applications or browsing private information without permission. They ascribed this risk to two types of intruders, outsiders and insiders. An insider is a person who may know the legitimate user and may have enough information about a legitimate user's patterns of behavior. \\

Muslukhov et al. provided the first empirical evidence that smartphone users consider insider threat to be important, underscoring the fact that such threats impact many smartphone users today. The authors developed new approaches to protect sensitive data in an unattended smartphone from unauthorized access by any individual who poses insider threat. They also proposed an adversarial model that describes capabilities and goals of both outsiders and insiders. Muslukhov et al. asked two relevant questions.
\begin{itemize}[\IEEEsetlabelwidth{Z}]
	\item Do users consider insiders to be a serious threat?
	\item Does unauthorized access by insiders occur in the real world?
\end{itemize}

Muslukhov et al. performed two studies. The first study interviewed 22 users, while the second study was an online  survey of 724 users. 
They find the following
\begin{itemize}[\IEEEsetlabelwidth{Z}]
	\item The majority of the users believe the insider risk is as critical as threats from outsiders.  
	\item More than 12\% of the users have faced unauthorized access of their data or applications on smartphones by insiders.\\
	more
	\item More than 9\% of subjects accessed smartphone without the owner's permission. \\ 
\end{itemize}

\subsubsection{General Security Concerns and Issues with Apps}

Chin et al. \cite{chin2012measuring} enumerated concerns related to smartphone security and privacy. The authors performed interviews with 60 participants. They also examined the reasons why individuals do not use certain security and privacy methods. They found that participants are concerned about privacy on their phones more than on their laptops. As a result, they were less likely to purchase or download apps that perform sensitive tasks (e.g., accessing health data) on their phones. Some of the users' apprehensions probably stem from misconceptions about the security of  the applications, as well as gaps in their understanding of the security of wireless connections versus security in end-to-end wired networks. To establish successful security indicators, the authors studied why users may decide to download certain applications. Users preferred to install more applications, like games and entertainment, on mobile devices than on laptops. The authors discovered that users found applications via browsing, advertisements, and recommendations by friends. They were likely to be less brand-conscious and more price-conscious while installing applications on their mobile devices. They also mostly ignored the applications' terms of service and policy agreements.\\

Based on their findings, Chin et al. suggested improvements that might promote smartphone security, helping users confidently harness the full potential of mobile applications.

\begin{itemize}[\IEEEsetlabelwidth{4)}]
	\item Education of users about the security features of various media types is necessary. In particular, highlighting the benefits of end-to-end encryption may go a long  way in helping to clear user misconceptions.
	\item Improvement of data backup, use of a lock mechanism, and removal of wiping services that are also topics require education.
	\item Implementation of new security indicators to increase user trust in the selection of applications is also essential. New security indicators in centralized smartphone application marketplaces could also help smartphone users identify trustworthy brands.
\end{itemize}

\subsubsection{\textbf{Importance of Security Concerns People Have}}

Muslukhov et al.\cite{muslukhov2012understanding} presented a qualitative user study to establish requirements necessary to protect valuable information for smartphone users. They believed that to provide security, it is necessary to have an understanding of how users classify sensitive data on smartphones. They interviewed 22 participants, asking questions such as the following: What type of data do users often store on their mobile phones? Where does valuable or sensitive data come from? What types of activities do users perform to ensure the accessibility, integrity and privacy of their data? The study concluded that when users choose to store sensitive data on their smartphones, they anticipate whether or not their data might be threatened. However, users did not consider specific details of possible risks. Users did not try to prevent probable threats, nor they did anything to ensure availability, confidentiality or integrity of information. Users did not use lock systems for their devices because they believe locking methods are ineffective. This point of view comes from the general popularity of Internet browsers, games and weather forecast applications, which do not require secure authorization. The survey also showed that users think that valuable data can be protected simply by saving information on local storage such as external hard drives.

\section{Lessons Learned, Open Problems and Future Trends}
\label{sec:LLOP}

The use of continuous authentication, based on behavioral biometrics is a promising area of study to enhance security and usability of smartphones. Our survey has analyzed a large number of prior studies that have used six types of behavioral biometric: handwaving, keystroke, touchscreen behaviors, gait, signatures, voice and behavioral profiling. These studies extract various features such as timestamps, pressure, finger size, touch location and gesture type and employ machine learning techniques. We investigate tens of existing approaches and compare published results from them. However, we believe that the results, in terms of accuracy and equal error rate, can be further improved if one is able to choose the most appropriate features and machine learning algorithms.\\

In this  section,  we discuss lessons learned from the studies presented in our survey, identify some problems that are appropriate for research in the future and discuss current and future trends.\\

\subsection{Lessons Learned}

Behavioral biometrics are considered promising for providing continuous authentication for consumers. This survey has discussed many methods that can be used to enhance smartphone security. \\

\subsubsection*{Evaluation Metrics}
One of the main main goals of incorporating behavioral biometrics is to offer an easy way to access mobile devices with few interruptions once the owner wants to use the device. With a wide variety of proposed approaches, there is need for a general and widely accepted metric to compare different approaches. Currently, there is a plethora of evaluation metrics, but none is accepted as a standard.\\

\subsubsection*{Using Machine Learning}
Some studies implement approaches using machine learning algorithms while others do not. Machine learning algorithms are well-suited  to generalize from past user behaviors to predict the future and as a result, are likely to be most appropriate for passive authentication.\\

\subsubsection*{Balancing security and usability}

One of the major goals in continuous authentication is balancing between security and usability to provide easy use of the device and offer high level of security at the same time. Studies such as Gascon et al. \cite{gasconcontinuous} and De Luca et al. \cite{de2012touch}, which have high value of FPR suffer from low usability causing the device to be locked continuously, which means the user has to validate each time he/she wants to use the device. This action is incompatible with the purposes of continuous authentication. Therefore, the FPR has to be reduced to avoid this issue.\\  

\subsubsection*{Dealing with legitimate but unusual behaviors}

Many studies such as \cite{zheng2012you} build on the assumption that a user's behavior is consistent and no abrupt changes happen over a short period of time. However, this assumption may not always be true. For example, a physical injury or a panic situation may lead to unexpected behavioral changes, which can lead to inconsistent reactions to the situation by the legitimate user. Therefore, proper handling of unexpected situations is absolute necessary.\\ 

\subsubsection*{Differentiation between legitimate user and insider attack} 

Differentiating between a legitimate user and an inside attacker is not easy, especially when someone may be able to realistically impersonate the legitimate user,by mimicking application usage and other behaviors of legitimate users . \\  

\subsection{\textbf{Open Problems}}

In this section, we discuss potential solutions to open problems that need to be overcome to further facilitate the use of behavioral biometrics. This discussion occurs within the context of smartphone security, particularly of authentication.\\

\subsubsection{Increasing Accuracy} 

 An open problem is that the accuracy of current studies has to improve.  We suggest that be increased by using accuracy combination of methods. For example, in the context of touchscreen usage, a mechanism to gauge the authenticity may use a variety of features, such as single-touch, multi-touch, touch movement, touch direction, touch pressure, and touch size. Other possible useful features may be fingertip pressure and the size and shape of the contact area between the fingertip and the touchscreen. \\    

\subsubsection{Controlled Environment}

 Most current studies gather and record data under laboratory conditions. The problem is that the results that are obtained in a lab may not accurately reflect what could occur in reality. Therefore, an approach to enhancing authentication should investigate the influence of external variables that would occur in realistic contexts (e.g., usage while walking or standing or in a vehicle) or other environmental situations that frequently occur within mobile contexts.\\

\subsubsection{Smartphone platform} 

 Most published methods have been tested on the Android platform primarily because of its popularity openness, and wide availability on smartphones from vendors around the world. These methods have ignored other platforms like iOS Windows Mobile and Symbian that also have built-in sensors, such as accelerometers and gyroscopes. Therefore, in order to insure wider availability of consistent authentication mechanism, the investigation on other platforms should be performed.\\

\subsubsection{Energy consumption}

 Another open problem is that most studies do not focus on the topic of resource consumption. Researchers know that this topic should be one of the main concerns when evaluating an approach to authentication. During an evaluation of how resources are used and energy is consumed, one must take certain issues into account, such as the number of features analyzed, the number of training profiles, the computational complexity of the approach, and usage of CPU, memory, and battery. These considerations must be an integral part of any viable authentication mechanism.\\

\subsubsection{Building Corpus} 
 Approaches to authentication that are based on what can be learned from machines suffer from a lack of sufficient data that are publicly available. A reason for this problem is that it is expensive to collect data over a number of days, weeks, or months, especially with large numbers of diverse subjects. Thus, the solution is to construct a corpus that has a large amount of appropriate data and can be used to obtain valid, interesting results. However, the construction of such a corpus must also overcome privacy, policy and regulatory hurdles that it is done legally and shared.\\
 
A study by Ngo et al. \cite{ngo2014largest} has built a large database for the study of sensor-based gaits, with 744 participants (355 females and 389 males) over the course of five days. In order to be valid for use with machine learning, other types of behavioral biometrics also require a large amount of publicly available data.\\

\subsubsection{Mobile Intrusion Detection Systems (IDS)}
 Detecting intrusions should be based on computationally efficient methods. Most investigators do not include a sufficient number of criteria when investigating a mobile IDS. Many aspects need to be included, such as implementing novel decision engines and reducing power consumption. A written checklist would help investigators remember all of the necessary criteria.\\

\subsubsection{Application Usage}
 
 The other problem involves not knowing how to distinguish among users. With millions of available apps in online markets, many users tend to use a regular selection of the same apps every day for specific purposes. In order to pinpoint which users are utilizing which apps, one needs to observe user behavior, including which apps they access and how and where they access them.\\

 \subsection{Future Trends}
 
With users accessing smartphones constantly, there is a need to offer a continuous authentication mechanism in order to reduce the frequency of entering PINs or drawing patterns. Continuous authentication can be based on two of behavioral biometrics: app usage and touch based. With widespread use of apps (the number of mobile app downloads expected to reach 268.69 billion in 2017 \cite{APD} that work on touchscreens (most current devices support touchscreen), using both for authentication sounds natural. \\

We believe that these types of biometrics when we are able to extract the most useful features and a device is being accessed can offer high level of security. Hence, the future of authentication of smartphones is likely to incorporate continuous authentication using app usage and touchscreen behavior to provide a constant high level of security while providing unfettered usability.\\

\section{Conclusion}
\label{sec:CON}
The growing number of users of smart device is resulting in an increasing amount of private information being stored inside each such devices. Numerous problems in security and privacy are constantly being raised. To resolve these issues, researchers have implemented many methods including continuous authentication approaches based on user behavior. This paper has discussed and compared a number of existing solutions from several perspectives. New methods must focus on multiple characteristics and secure against a variety of attacks, while making the security system easy to use and adapted to each owner.\\ 

In addition to the methods discussed above, a promising approach may be to measure user behavior in terms of application usage. Each smartphone contains applications which can be used for various purposes. Therefore, making continuous authentication based on application usage can be one way to enhance security and privacy. For example, applications can be categorized into social applications such as Twitter, Facebook and Google+; media applications such as those related to photos, camera and video; chatting applications such as WhatsApp, Snapchat and BBM; and transaction applications such as bank and credit card applications. Quantifying how, when and for how much duration these applications are accessed by specific users may help an implicit authentication system learn manners of fine-grained use and differentiate between authorized and non-authorized persons. Building a corpus for different patterns of app usage with a large number subjects over a number of days will be an excellent way to contribute to the field.\\

\section*{Acknowledgment}
The authors wish to thank the anonymous reviewers for their helpful and valuable comments that greatly contributed to improving the manuscript.\\

\bibliographystyle{IEEEtran}
\bibliography{survey}

\begin{IEEEbiography}[{\includegraphics[width=1in,height=1.5in,clip,keepaspectratio]{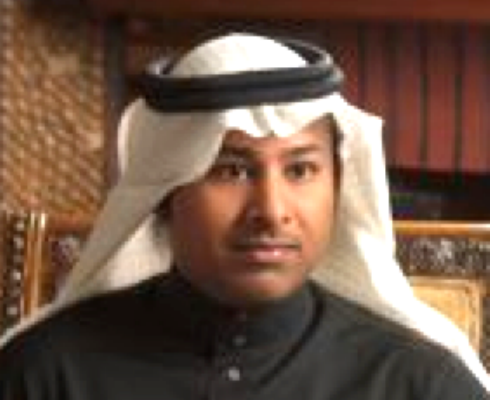}}]{Abdulaziz Alzubaidi} received his B.S. degree in Computer Science from University College, King Abdulaziz University, Saudi Arabia, in 2001. He received his MS degree in Computer Science from Jordan University, Jordan in 2009. He is currently pursuing the Ph.D. degree in Computer Security degree with the Department of Computer Science, University of Colorado at Colorado Springs, Colorado Springs, CO, USA. His research interests include biometrics, human activity recognition and smartphone security. His current work concentrates on continuous authentication of smartphone users using behavioral biometrics.
\end{IEEEbiography} 

\begin{IEEEbiography}
	[{\includegraphics[width=1in,height=1.25in,clip,keepaspectratio]{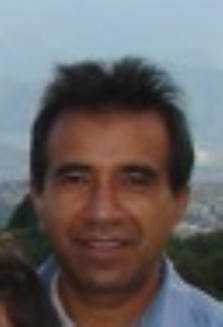}}]{Jugal Kalita} is a professor of Computer Science at The University of Colorado at Colorado Springs. He received his Bachelor of Technology degree from the Indian Institute of Technology in Kharagpur, India, his Master of Science degree from the University of Saskatchewan, Canada, and a Master of Science and PhD from the University of Pennsylvania. His research interests are in machine learning and its applications to areas such are natural language processing, intrusion detection and bioinformatics. He is the author of 160 papers in reputed conferences and journals. 
\end{IEEEbiography}

\end{document}